\documentclass[twocolumn,prd,nofootinbib,aps,floats,floatfix,amsmath,amssymb,longbibliography,
secnumarabic]{revtex4}
\usepackage[final]{graphicx}
\usepackage[utf8]{inputenc}
\usepackage[breaklinks,colorlinks,urlcolor=blue,citecolor=blue,linkcolor=magenta]{hyperref}
\usepackage{xcolor}
\usepackage{pifont}
\usepackage{verbatim}
\newcommand{\cmark}{\ding{51}}
\newcommand{\xmark}{\ding{55}}

\def\sfrac#1#2{{\textstyle{#1\over #2}}}
\newcommand{\be}{\begin{equation}}
\newcommand{\ee}{\end{equation}}
\newcommand{\ba}{\begin{array}}
\newcommand{\ea}{\end{array}}
\newcommand{\bea}{\begin{eqnarray}}
\newcommand{\eea}{\end{eqnarray}}
\newcommand{\sss}{\scriptscriptstyle}

\newcommand{\tr}{{\rm tr}}

\newcommand{\nn}{\nonumber}
\renewcommand{\L}{{\sss L}}
\newcommand{\R}{{\sss R}}
\newcommand{\T}{{\sss T}}
\newcommand{\E}{{\sss E}}
\newcommand{\N}{{\sss N}}

\newcommand{\diff}{\mathrm{{d}}}

\begin{document}

\title{Gauging lepton flavor SU(3) for the muon $g-2$}
\author{Gonzalo Alonso-\'Alvarez}
\author{James M.\ Cline}

\affiliation{McGill University, Department of Physics, 3600 University St.,
Montr\'eal, QC H3A2T8 Canada}

\begin{abstract}

Gauging a specific difference of lepton numbers such as $L_{\mu}-L_{\tau}$ is a popular model-building option, which gives rise to economical explanations for the muon anomalous magnetic moment.
However, this choice of gauge group seems rather arbitrary, and additional physics is required to reproduce the observed neutrino masses and mixings.
We address these shortcomings by embedding $L_{\mu}-L_{\tau}$ in the  vectorial SU(3) gauge symmetry of lepton flavor.
The vacuum expectation values (VEVs) of scalar fields in the fundamental, six-dimensional and adjoint representations allow for phenomenologically viable lepton and gauge boson masses.
The octet scalar gives rise to charged lepton masses, and together with the triplet scalar generates masses for all the leptophilic gauge bosons except for the $L_{\mu}-L_{\tau}$ one.
The latter gets its smaller mass from the sextet VEVs, which also generate the neutrino masses, and are determined up to an overall scaling by the observed  masses and mixings.
The model predicts three heavy neutral leptons at the GeV-TeV scale as well as vectorlike charged lepton partners; it
requires the mass of the lightest active neutrino to
exceed $10^{-4}$\,eV, and it naturally provides a resolution of the Cabibbo angle anomaly. 

\end{abstract}

\maketitle

%%%%%%%%%%%%%%%%%%%%%%%%%%%%%%%%%%%%%%%%%%%%%%%%%%%%%%%%%%%%%%%%%%%%%%
%%%%%%%%%%%%%%%%%%%%%%%%%%%%%%%%%%%%%%%%%%%%%%%%%%%%%%%%%%%%%%%%%%%%%%

\section{Introduction}

Leptophilic gauge bosons are popular candidates for physics beyond the Standard Model (SM).
As opposed to new gauge bosons interacting with quarks, which are strongly constrained by LHC searches~\cite{ATLAS:2017eqx,CMS:2018mgb,ATLAS:2019erb,CMS:2021ctt}, those coupling solely to leptons are only subject to LEP constraints~\cite{ALEPH:2013dgf} and can therefore exist below the TeV scale.
As a consequence, these kind of $Z'$ bosons can have important phenomenological consequences for a plethora of particle physics experiments and observations~\cite{Bauer:2018onh}. 

In view of recent experimental developments, $Z'$ bosons interacting with muons are of particular interest.
Most importantly, they can economically solve the long-standing discrepancy between the SM prediction~\cite{Aoyama:2020ynm} and the observed values~\cite{Muong-2:2006rrc,Muong-2:2021vma} of the muon anomalous magnetic moment, as has been shown in~\cite{Foot:1994vd,Gninenko:2001hx,Baek:2001kca,Murakami:2001cs,Fayet:2007ua,Pospelov:2008zw,Ma:2001md,Heeck:2011wj,Carone:2013uh,Harigaya:2013twa,Davoudiasl:2014kua,Altmannshofer:2014cfa,Tomar:2014rya,Lee:2014tba,Allanach:2015gkd,Heeck:2016xkh,Patra:2016shz,Altmannshofer:2016brv,Iguro:2020rby,Holst:2021lzm,Hapitas:2021ilr}.
Slight deviations from SM expectations in lepton flavor universality (LFU) ratios~\cite{HFLAV:2019otj} and an observed deficit of unitarity in the first row of the CKM matrix~\cite{Belfatto:2019swo,Grossman:2019bzp,Shiells:2020fqp,Seng:2020wjq} may also be explained by a new muonic force.
Finally, accumulating evidence for lepton-universality violation in $b\rightarrow s\ell^+\ell^-$ transitions (see e.g.~\cite{Geng:2021nhg} and references therein) serves as a further motivation to study this kind of new physics (NP), although in this case coupling to quarks also need to be invoked.
A global analysis of leptophilic gauge bosons was recently performed in~\cite{Buras:2021btx}.

In this context, the most popular phenomenological model is that of a gauge boson coupling to the $L_\mu - L_\tau$ lepton flavor combination~\cite{He:1990pn,Foot:1990mn,He:1991qd}.
Gauged $U(1)_{L_\mu-L_\tau}$ models are distinguished by being free from gauge anomalies (as are other differences in baryon and/or lepton flavor numbers), and from the stringent experimental constraints on $Z'$ couplings to electrons.

One difficulty is that gauging $L_\mu-L_\tau$ symmetry prevents the usual interactions that generate the masses and mixings of neutrinos through the dimension-5 Weinberg operator.
This means that a fully consistent model requires additional new physics to reproduce the observed values~\cite{Esteban:2020cvm} of the PMNS~\cite{Maki:1962mu,Pontecorvo:1967fh} matrix. 
Some proposals in this direction include extra Higgs doublets~\cite{Ma:2001md}, soft-breaking terms~\cite{Bell:2000vh,Choubey:2004hn}, and right-handed neutrinos~\cite{Binetruy:1996cs,Heeck:2011wj,Asai:2017ryy,Araki:2019rmw}.
The common element of all these models is the presence of (often multiple) U(1)$_{L_\mu-L_\tau}$ symmetry-breaking scalars that can significantly complicate the minimal proposal.
From an aesthetic point of view, it also seems rather arbitrary that only the $L_\mu-L_\tau$ difference should be gauged, considering that the SM treats all generations on an equal footing from a structural perspective.

To address these shortcomings, in this paper we propose that SU(3)$_\ell$ of lepton flavor is gauged in a vectorial fashion.
In this setup, the $L_\mu-L_\tau$ gauge boson need only be one of eight new $Z'$ states, all of which are leptophilic. Previous studies have gauged lepton and quark flavors together using horizontal SU(3) family symmetry, 
along with additional gauge or discrete symmetries~\cite{Berezhiani:1983hm,King:2001uz,King:2003rf,Alonso:2017bff}, or flavor symmetries for left- and right-handed leptons separately~\cite{Alonso:2016onw}. We are not aware of previous literature treating the  possibility of a single vectorial SU(3)$_\ell$ for lepton flavor. 

In the following, we identify the additional particle content needed to spontaneously break SU(3)$_\ell$ to engender the observed lepton and neutrino masses, which would otherwise be forced to be flavor universal.
Further guided by experimental observations, we focus on scenarios where the $L_\mu-L_\tau$ gauge boson is lighter than the other seven and can thus address the $(g-2)_\mu$ tension. 
For these purposes, we find it sufficient to add a triplet of vectorlike heavy leptons $E_i$ and three scalars in the $3$, $6$ and $8$ representations, which we denote by $\Phi_3$, $\Phi_6$, and $\Phi_8$, respectively.
The usual triplet of right-handed neutrinos
$N_i$ is also present.
The vacuum expectation values (VEVs) of the scalars give rise to the required lepton and gauge boson masses.
This particle content is summarized in Table \ref{tab1}.

A modest hierarchy in the scalar VEVs, $\langle\Phi_3\rangle \sim\langle\Phi_8\rangle \gg \langle\Phi_6\rangle$ is needed to get the desired spectrum of leptophilic gauge boson masses.
The gauge boson coupling to the $L_\mu-L_\tau$ current is lighter than the rest, which allows it to solve the muon $(g-2)$ anomaly in a way consistent with all other constraints~\cite{Holst:2021lzm,Hapitas:2021ilr}.
At the same time, the Cabibbo angle anomaly (CAA), a 3-5\,$\sigma$ tension in the unitarity of the top row of the CKM matrix~\cite{Belfatto:2019swo,Grossman:2019bzp,Shiells:2020fqp,Seng:2020wjq}, can be resolved by new contributions to $\mu\to e\nu\bar\nu$ decay from two of the heavier new gauge bosons. 
A mild 2$\sigma$ hint of nonuniversality in $\tau$ lepton decays~\cite{HFLAV:2019otj} can be similarly ameliorated.

The $\Phi_6$ VEVs further determine the mass matrix of the right-handed neutrinos, and thereby the pattern of light neutrino masses and mixings via a type I seesaw mechanism~\cite{Minkowski:1977sc,Gell-Mann:1979vob,Yanagida:1979as,Glashow:1979nm,Mohapatra:1979ia,Weinberg:1979sa,Witten:1979nr}.
A striking prediction of our scenario is that the lightest active neutrino mass must exceed $\sim 0.1\,\mathrm{meV}$.
We find the right-handed neutrino masses to lie in the GeV to TeV range and to be inversely proportional to the light neutrino masses.
They are therefore heavy neutral leptons
(HNLs), whose mixings with the active states exactly match the flavor composition of the light neutrino mass eigenstates and are therefore completely determined by the PMNS matrix.  We show  that the mixings of the HNLs with active neutrinos can be in a range that is relevant for affecting the primordial abundances of light elements.

The paper is structured as follows. In Section \ref{sect:model} we introduce the particle content of the model and the resulting spectra of charged leptons, heavy vectorlike leptons, light neutrinos, and
new gauge bosons.  Constraints on the gauge bosons
from flavor-sensitive processes are discussed in
Section \ref{sect:pheno}, including the muon $(g-2)$, LEP di-lepton searches, the Cabibbo angle anomaly, and lepton flavor universality limits.  The detailed properties of the
heavy neutral leptons, and their phenomenological implications, 
are discussed in Section~\ref{sec:HNL}.
Conclusions are given in Section~\ref{sect:conc}, and a brief discussion of the challenges associated with the construction of a potential in the scalar sector that can give rise to the desired pattern of VEVs is presented in Appendix~\ref{app:VEVs}.

%%%%%%%%%%%%%%%%%%%%%%%%%%%%%%%%%%%%%%%%%%%%%%%%%%%%%%%%%%%%%%%%%%%%%%
%%%%%%%%%%%%%%%%%%%%%%%%%%%%%%%%%%%%%%%%%%%%%%%%%%%%%%%%%%%%%%%%%%%%%%

\section{Model and mass spectra}
\label{sect:model}

Our starting point is the gauging of the leptonic SU(3) flavor symmetry, under which both left- and right-handed lepton fields transform simultaneously.
Clearly, our world does not exactly respect such a symmetry, requiring it to be spontaneously broken  at some scale.
For this purpose, we introduce three real scalar fields $\Phi_3$, $\Phi_6$, and $\Phi_8$ in the fundamental, symmetric two-index, and adjoint representations, respectively.
Although $\Phi_6$ by itself would be sufficient to fully
break SU(3)$_\ell$,  this is the minimal scalar sector that we have identified as being phenomenologically viable,\footnote{Other possibilities may exist; it is not
our purpose to exhaust them but rather to construct one instance of a working model.} as we will explain.

Clear observational evidence for breaking of SU(3)$_\ell$ is provided by the hierarchy of lepton masses.
Indeed, with only the SM particle content, the gauge-invariant Higgs coupling $\bar L^i H e_i$ (where $i$ is the SU(3)$_\ell$ index) only allows for a universal charged lepton mass.
To obtain the observed mass splittings through spontaneous breaking of the gauge symmetry in a renormalizable way, we add a triplet of vectorlike charged lepton partners, $E_i$.
Additionally, to generate neutrino masses, a right-handed neutrino triplet $N_i$
is included, which completes the particle content of the model.
The list of particles and gauge charges are shown in Table~\ref{tab1}.

\begin{table}[t]\centering
\setlength\tabcolsep{4pt}
\def\arraystretch{1.2}
\begin{tabular}{|c|c|c|c|}\hline
 & SU(3)$_\ell$ & SU(2)$_L$ & U(1)$_y$\\
\hline
$L_i$ & 3 & 2 & $-1/2$ \\
$e^{c,i}$ & $\bar 3$ & 1 & $+1$ \\
\hline
$E_{\L,i}$ & 3 & 1 & $-1$ \\
$E_R^{c,i}$ & $\bar 3$ & 1 & $+1$ \\
$N^{c,i}$ & $\bar 3$ & 1 & 0 \\
\hline
$\Phi_{3,i}$ & 3 & 1 & 0 \\
$\Phi_{6,ij}$ & 6 & 1 & 0 \\
$\Phi^i_{8,j}$ & 8 & 1 & 0\\
\hline
\end{tabular}
\caption{Field content of the gauged SU(3)$_\ell$ model along with the corresponding gauge charges. The fermionic fields are all listed as left-handed Weyl states.  $L_i$ and $e_i$ denote the SM lepton doublets
and charged singlet leptons, respectively.}
\label{tab1}
\end{table}

The addition of the right-handed neutrino triplet $N_i$ makes SU(3)$_\ell$ anomaly free, since all leptons now come in chiral pairs. 
Mixed anomalies involving non-abelian factors likewise cancel because of the tracelessness of the generators, and the SU(3)$_\ell^2\times$U(1)$_y$ one vanishes due to the hypercharges of the SU(3)$_\ell$ triplets adding up to zero.

With the field content listed in Table~\ref{tab1}, the most general renormalizable Lagrangian that can be constructed (omitting kinetic terms) is
\bea\label{eq:Lagrangian}
{\cal L} = 
    &-& \bar E_\L^i(\mu_{\E\E}\delta^j_i + y_{\E\E}\Phi^j_{8,i}) E_{\R,j}\nn\\
    &-& \bar E_\L^i(\mu_{\E e}\delta_i^j + y_{\E e}\Phi_{8,i}^j)e_j \nn\\
    &-&\bar L^i H(y_{\L e} e_i + y_{\L\E} E_{\R,i}) -	y_{\L\N} \bar L^i \tilde H N_i\nn\\
	&-& {y_\N\over 2} \bar N^i\Phi_{6,ij}N^{c,j} + {\rm h.c.} \nn \\
	&-& V(H,\, \Phi_3,\, \Phi_6,\, \Phi_8),
\eea
where $V$ denotes the scalar potential, $\mu_i$ are
constants with dimensions of mass, and $y_i$ are dimensionless couplings.
In the following sections we describe the consequences of nonvanishing vacuum expectation values for the scalar fields.
The required values of the VEVs are constructed in a bottom-up fashion, making use of experimental measurements and constraints as guiding principles.
Obtaining the desired symmetry breaking pattern from a
specific scalar potential is likely to be a challenging task, beyond the scope of the present work.
We however take some preliminary steps in this direction 
in Appendix~\ref{app:VEVs}.

In the following, it will be convenient to order the gauge indices associated with the lepton flavors as $(1,2,3) = (\tau,\mu,e)$.
This allows the $\mu-\tau$ gauge boson to be associated with the third generator $T_3$ of SU(3)$_\ell$ (using the standard form of the Gell-Mann matrices) instead of a linear combination of $T_3$ and $T_8$.

%%%%%%%%%%%%%%%%%%%%%%%%%%%%%%%%%%%%%%%%%%%%%%%%%%%%%%%%%%%%%%%%%%%%%%

\subsection{Charged lepton masses}

\begin{figure}[t]
\centering
\includegraphics[width=0.4\textwidth]{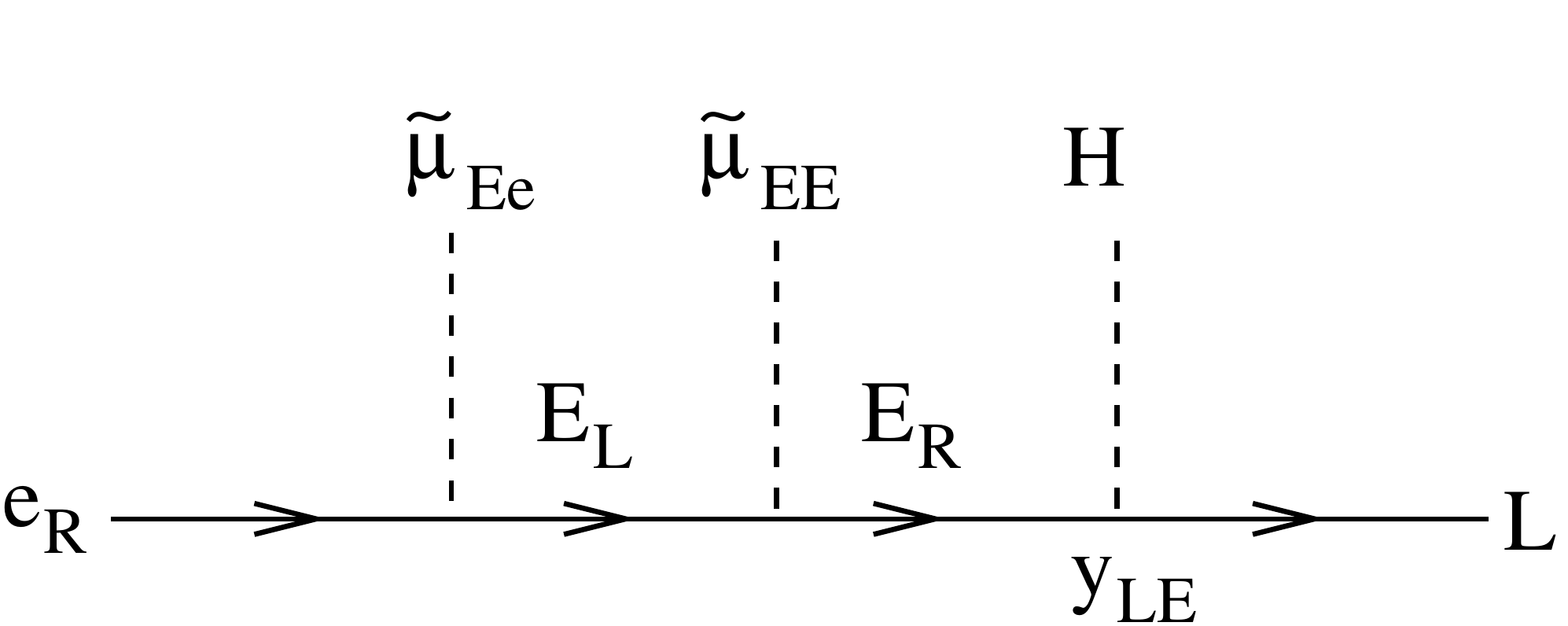}
\caption{Extra contribution to the charged lepton masses from integrating out the heavy vectorlike lepton, that splits them relative to the flavor-independent mass  $y_{\L e}\bar v$.}
 \label{fig:diag}
\end{figure}

In the absence of spontaneous breaking of SU(3)$_\ell$, charged leptons get a universal mass from electroweak
symmetry breaking.  We will use the $\Phi_8$ scalar field VEV, in conjunction with the vectorlike lepton partners $E_i$, 
to split masses between the different lepton generations.\footnote{Nonzero $\mu_{\E\E}$ or $\mu_{\E e}$ values in Eq.~\eqref{eq:Lagrangian} are also essential, since in the $\mu_{\E\E},\,\mu_{\E e}\rightarrow 0$ limit the charged lepton masses become universal again.}\ \ 
The Lagrangian~\eqref{eq:Lagrangian} gives rise to
mass mixing between the SM leptons and the heavy $E_i$
partners.
For simplicity, we take $\langle\Phi_8\rangle$ to be diagonal, in which case the $6\times 6$ Dirac mass matrix for the charged leptons becomes block diagonal
(no flavor mixing),
with three blocks of the form
\be \label{eq:Dirac_mass_matrix}
    (\bar e_\L^i,\, \bar E_\L^i)\left({y_{\L e}\bar v\atop \tilde \mu^i_{\E e}}\,{ y_{\L\E} \bar v\atop \tilde \mu^i_{\E\E}}
    \right)\left({e_{\R,i}\atop E_{\R,i}}\right),
\ee
where $\bar v \cong 174\,$GeV is the complex Higgs VEV and we define
\bea
    \tilde \mu^i_{\E e} &=& \mu_{\E e} + y_{\E e}\langle\Phi_8\rangle^i_i\,,\nn\\
    \tilde \mu^i_{\E\E} &=& \mu_{\E\E} + y_{\E\E}\langle\Phi_8\rangle^i_i\,.
    \label{tildedef}
\eea
Diagonalizing each of the $2\times2$ blocks separately under the assumption $y_{\L e}\bar{v},\,y_{\L\E}\bar{v}\ll \tilde\mu^i_{\E\E}$ or $\tilde\mu^i_{\E e}$ leads to three heavy and three light mass eigenstates with masses
\bea
    (m_\E^i)^2 &=& (\tilde\mu^i_{\E e})^2 + (\tilde\mu^i_{\E\E})^2,\nn\\
    (m_e^i)^2 &=& \bar{v}^2\,\frac{(y_{\L e}\tilde\mu_{
    \E\E}^i - y_{\L\E}\tilde\mu_{
    \E e}^i)^2}{(\tilde\mu^i_{\E e})^2 + (\tilde\mu^i_{\E\E})^2}.\label{eq:charged_lepton_masses}
\eea
The contributions to the light charged lepton masses in the perturbative limit are depicted in Fig.\ \ref{fig:diag}. 
In the limit of large $\mu_{\E\E}$ or $\mu_{\E e}$, the heavy lepton effects do not decouple; lepton masses generally continue to be split by the heavy-$E_i$ effects even for arbitrarily heavy $E_i$.

Since we assume that $\langle\Phi_8\rangle$ is diagonal, mixing with the heavy states does not cause any flavor-violating effects in the charged lepton sector.
However, the mixing of the light states with the heavy ones is constrained by electroweak precision data (EWPD).
Diagonalizing  the  mass matrix in Eq.~\eqref{eq:Dirac_mass_matrix} yields mass eigenstates $(e^i_{\L,\R},\,E^i_{\L,\R})'$, given by
\be
\left({e^i_{\L,\R}\atop  E^i_{\L,\R}}\right) = 
\left({c_{\theta^i_{\L,\R}}\atop -s_{\theta^i_{\L,\R}}}\,{s_{\theta^i_{\L,\R}}\atop
\phantom{-}c_{\theta^i_{\L,\R}}}\right)\left( e_{\L,\R}^i \atop E_{\L,\R}^i \right)'\,.
\ee
To second order in $\bar{v}/\tilde\mu$, the mixing angles are given by
\bea
\tan 2\theta^i_\L &=& \frac{2\bar v\,(y_{\L e}\tilde \mu^i_{\E e} + y_{\L\E}\tilde \mu^i_{\E\E})}{(\tilde \mu^i_{\E e})^2 + (\tilde \mu^i_{\E\E})^2}\,,\nn\\
\tan 2\theta^i_\R &=& 2 \frac{\tilde\mu^i_{\E\E} \tilde\mu^i_{\E e} + \bar{v}^2 y_{\L\E}y_{\L e}}{(\tilde\mu^i_{\E\E})^2 - (\tilde\mu^i_{\E e})^2} \,.\label{eq:charged_lepton_mixings}
\eea
In the limit where $\tilde \mu^i_{\E\E}$ or $\tilde\mu_{\E e}^i\gg \bar v$, the angle $\theta^i_\L$ is suppressed, but $\theta^i_\R$ can be relatively large.
EWPD constraints require that $\theta_{\R}(\tau,\mu,e) \lesssim (0.03,\,0.02,\,0.02)$~\cite{delAguila:2008pw}.\footnote{The precise upper limits are model-dependent; we have chosen the strongest limits corresponding to models that give dominant right-handed mixing.}\ \ 
To sufficiently suppress $\theta_\R^i$, there must be a hierarchy $\tilde \mu^i_{\E e} \ll \tilde \mu^i_{\E\E}$ or $\tilde \mu^i_{\E \E} \ll \tilde \mu^i_{\E e}$ for each value of $i$. 

To give a concrete example, consider the 
case $\tilde \mu^i_{\E e} \lesssim 10^{-2}\, \tilde \mu^i_{\E\E}$; the derivation for the opposite possibility is completely analogous.
In this limit, the masses of the heavy charged leptons are simply
\begin{equation}
	m^i_{\E} \simeq \tilde \mu^i_{\E\E}\,,
\end{equation}
while the mass of the SM charged leptons becomes
\begin{equation}
	m^i_e \cong \bar{v} \left( y_{\L e} - y_{\L\E} \frac{\tilde \mu^i_{\E e}}{\tilde \mu^i_{\E\E}} \right).
\end{equation}
Taking  for simplicity $y_{\L\E}=1$ and $y_{\L e} = 0$ ($y_{\L e}\lesssim 10^{-6} y_{\L\E}$ is the sufficient condition), the ratios
\begin{equation}\label{eq:mu_tilde_ratios}
	\frac{\tilde \mu^i_{\E e}}{\tilde \mu^i_{\E\E}} \simeq - \frac{m_e^i}{\bar{v}} \sim (10^{-2},\, 10^{-3},\, 10^{-6})
\end{equation}
are determined by the measured charged lepton masses.
Using Eq.~\eqref{tildedef} and taking into account the fact that $\Phi_8$ is traceless, 
one sees that achieving the hierarchies in Eq.~\eqref{eq:mu_tilde_ratios} requires some cancellations between terms in the numerator or denominator.
For example,  take  $\langle\Phi_8\rangle = 
\mathrm{diag}(-\mu_{\E\E},\,\mu_{\E\E},\,0)$.
Then  Eq.~\eqref{eq:mu_tilde_ratios} is realized by choosing
\begin{align}
    \mu_{\E e} &= \frac{m_e}{\bar{v}} \, \mu_{\E\E} \simeq 3\times 10^{-6}\, \mu_{\E\E} \,,\nn\\
    y_{\E\E} &= \frac{m_\tau+m_\mu}{m_\tau-m_\mu} \simeq 1.1\,,\nn\\
        y_{\E e} &= \frac{2}{\bar{v}} \frac{m_\tau m_\mu}{m_\tau-m_\mu}\simeq 1.3\times 10^{-3}\,.\label{eq:BM_parameters}
\end{align}
The required order-of-magnitude enhancement of $m_\tau$ relative to $m_\mu$ is achieved by a $\sim 10\%$ cancellation in the denominator of Eq.~\eqref{eq:mu_tilde_ratios} for $i=\tau$, while the smallness of $m_e$ is due to $\mu_{\E e}\ll
\mu_{\E\E}$ and $\langle\Phi_8\rangle_e^e = 0$.

Having fixed the charged lepton masses, one is still free to choose $\mu_{\E\E}$ and therefore the absolute mass scale of the charged lepton partners.
If they are sufficiently light, they may be directly accessible at the Large Hadron Collider (LHC).
They can be pair-produced by the Drell-Yan process, and their main decays are $E_i\to L_i + h$ via the
$y_{\L\E}$ interaction.  Ref.\ \cite{Kumar:2015tna} showed that electroweak-singlet vectorlike leptons (VLL's) are very difficult to observe at the LHC;  indeed, existing 
constraints \cite{ATLAS:2015qoy,CMS:2019hsm} have focused on the search for doublet VLL's.  
VLL models similar to ours were studied in 
Ref.\ \cite{Bell:2019mbn}, where current LHC limits were found to be as low as  $m_\E^i \gtrsim 300\,$GeV. This leaves a large range of allowed VLL masses for future discovery, consistent with our requirement $\langle\Phi_8\rangle\gtrsim 1$~TeV, that will be derived in Section~\ref{sect:pheno}.

In the following sections, we adopt the parameter values described above as our principal benchmark model (BM1).
However, other possibilities for fixing the constants in the Lagrangian~\eqref{eq:Lagrangian} exist.
There are eight free parameters: two dimensionful scales $\mu_{\E\E}$ and $\mu_{\E e}$, four Yukawa couplings $y_{\E\E}$, $y_{\E e}$, $y_{\L\E}$, and $y_{\L e}$, and two independent entries of $\langle\Phi_8\rangle$.
Reproducing the observed charged lepton masses imposes three constraints through Eq.~\eqref{eq:charged_lepton_masses}.
In addition, the mixings in Eq.~\eqref{eq:charged_lepton_mixings} must be sufficiently small.
In order to explore the parameter space and find other viable models, we employed a Markov Chain Monte Carlo method.
Two further examples of parameters that satisfy all the phenomenological requirements but differ qualitatively from BM1 are given in Table \ref{tab3}.  

In the second example (BM2), also expressed in terms of the unconstrained parameter $\mu_{\E\E}$, the heavy lepton masses are all approximately $m_\E^i\cong \mu_{\E\E}$.
Satisfying the EWPD bounds on mixing leads to the largest Yukawa coupling, $y_{\L\E} = 3.6$, being close to unitarity limits.
In this example, since $\mu_{\E\E}\gg\langle\Phi_8\rangle$, the heavy leptons are far above the TeV scale and therefore not directly accessible in current collider experiments.

In the third example (BM3) and similarly to BM1, the VLL are of order $\langle\Phi_8\rangle$, making them potentially accessible at the LHC.
In this case (as in BM1), all Yukawa couplings are well below perturbative unitarity bounds.  A qualitatively different feature of BM3 is that $\tilde\mu_{\E\E}^\tau \ll
\tilde\mu_{\E e}^\tau$, while in the others 
$\tilde\mu_{\E\E}^i \gg
\tilde\mu_{\E e}^i$ for all flavors.
Moreover, the VVL-charged lepton mixing angles fall well below the experimental limits in this model.

\begin{table}[t]\centering
\setlength\tabcolsep{6pt}
\def\arraystretch{1.4}
\begin{tabular}{|c| c c c|}\cline{2-4}
\multicolumn{1}{c|}{} & BM1 & BM2 & \multicolumn{1}{c|}{BM3} \\ \cline{1-4}
\multicolumn{4}{|c|}{Input parameters} \\ \hline
$\phi_\mu$ & $\mu_{\E\E}$ & $0.014\,\mu_{\E\E}$ & $-3.0\,\mu_{\E\E}$ \\ \hline
$\phi_e$ & $0$ & $0.016\,\mu_{\E\E}$ & $-3.2\,\mu_{\E\E}$ \\ \hline
$y_{\L\E}$ & $1$ & $3.6$ & $-0.011$ \\ \hline
\multicolumn{4}{|c|}{Derived parameters} \\ \hline
$\mu_{\E e}$ & $3\times 10^{-6}\,\mu_{\E\E}$ & $0.023\,\mu_{\E\E}$ & $1.55\,\mu_{\E\E}$ \\ \hline
$y_{\E\E}$ & $1.1$ & $0.32$ & $0.16$ \\ \hline
$y_{\E e}$ & $1.3\times 10^{-3}$ & $-0.055$ & $-0.48$ \\ \hline
$y_{\L e}$ & $\lesssim 10^{-6}$ & $0.56$ & $-8.5\times 10^{-4}$ \\ \hline
\multicolumn{4}{|c|}{Phenomenological quantities} \\ \hline
$m^i_E$ & $(1,\,1,\,1)\,\mu_{\E\E}$ & $(1,\,1,\,1)\,\mu_{\E\E}$ & $(4.5,\,1.5,\,1.5)\,\mu_{\E\E}$ \\ \hline
$\theta_{\R}^{\tau}$ & $1\times 10^{-2}$ & $0.021$ & $-1\times 10^{-3}$ \\ \hline
$\theta_{\R}^{\mu}$ & $6\times 10^{-4}$ & $5.6\times 10^{-4}$ & $6\times 10^{-3}$ \\ \hline
$\theta_{\R}^e$ & $3\times 10^{-6}$ & $0.024$ & $6\times 10^{-3}$ \\ \hline
\end{tabular}
\caption{Input parameters, derived parameters and quantities relevant for phenomenology for three benchmark models that can reproduce the observed charged lepton properties.
The mixing angles $\theta_\R^i$ between right-handed charged leptons and heavy vectorlike leptons of mass $m_E^i$ are consistent with EWPD constraints~\cite{delAguila:2008pw}.
The traceless octet VEV is parametrized as $\langle\Phi_8\rangle = \mathrm{diag}(-\phi_\mu-\phi_e,\,\phi_\mu,\,\phi_e)$.
The dimensionful parameter $\mu_{\E\E}$, which controls the scale of the octet VEV and the VLL masses, is left as freely adjustable parameter.
}
\label{tab3}
\end{table}

%%%%%%%%%%%%%%%%%%%%%%%%%%%%%%%%%%%%%%%%%%%%%%%%%%%%%%%%%%%%%%%%%%%%%%

\subsection{Neutrino masses}

The right-handed neutrinos $N_i$ get a Majorana mass matrix from the
sextet VEVs, $M_N = y_\N\langle\Phi_6\rangle$, while the Dirac
neutrino mass matrix is proportional to the unit matrix, with 
coefficient $m_D = y_{\L\N}\bar v$.  The light neutrino mass matrix is
therefore $m_\nu = -m_D^2 M_N^{-1}$ from the seesaw mechanism, and we can solve for the sextet
VEVs in terms of the PMNS mixing matrix $U_{\rm\sss PMNS}$ and the light neutrino
mass eigenvalues $D = {\rm diag}(m_1,m_2,m_3)$:
\be\label{eq:sextet_VEV}
	\langle\Phi_6\rangle = -{m_D^2\over y_\N}\,U_{\rm\sss PMNS}\, D^{-1}\, U_{\rm\sss PMNS}^\T\,.
\ee
Recall that we have interchanged the first and third rows of $U_{\rm\sss PMNS}$ in 
our ordering convention for the lepton flavors.  This allows us
to refer to the $L_{\mu-\tau}$ gauge boson as the one associated with
the Gell-Mann matrix $\lambda_3$.
A quantitative analysis of the right-handed neutrino mass and mixing spectrum is deferred to Sec.~\ref{sec:HNL}.

%%%%%%%%%%%%%%%%%%%%%%%%%%%%%%%%%%%%%%%%%%%%%%%%%%%%%%%%%%%%%%%%%%%%%%

\subsection{Gauge boson masses}

The mass-squared matrix of gauge bosons is given by
\bea
\label{gbmasses}
	{M^2_{ab}\over g'^2} &=& \sfrac14\Phi_3^\dagger \{\lambda^a,\lambda^b\} \Phi_3 
	- \sfrac12\tr\left([\Phi_8,\lambda^a][\Phi_8,\lambda^b]\right)\nn\\
	&+& 
\sfrac12\tr\left([\Phi_6^\dagger\lambda^a + 
\lambda^{a*}\Phi_6^\dagger][\Phi_6\lambda^{b*} + \lambda^b\Phi_6]\right)\,,
\eea
where the VEVs of the scalar fields are understood.
In the approximation that $\langle\Phi_6\rangle=0$, while
$\langle\Phi_3\rangle\sim \langle\Phi_8\rangle$, all gauge bosons get masses except for $A_3$,
which corresponds to the $L_{\mu-\tau}$ gauge boson in our numbering
scheme.  Thus $A_3$ gets its mass solely from $\langle\Phi_6\rangle$.
The overall magnitude of $\langle\Phi_6\rangle$ is determined by the 
smallest light neutrino mass $m_{\nu_1}$; hence one prediction of the model
is that $m_{\nu_1}$ cannot be arbitrarily small.

The various contributions of the scalar VEVs to the mass eigenvalues $M_i^2/g'^2$ are summarized in 
Table~\ref{tab:gauge_boson_masses}.
Before turning on the $\Phi_6$ VEV, the gauge boson mass matrix is diagonal, and the eigenvalues can be labelled with the number of the corresponding generator.

We first consider the contribution of the $\Phi_8$ VEV. Following the discussion in the previous section, in  benchmark scenario BM1 it takes the form
\be \label{eq:Phi8_VEV}
    \langle\Phi_8\rangle \equiv \phi_8\,{\rm diag}(-1,\,1,\,0)\,,
\ee
with $\phi_8 = \mu_{\E\E}$.\footnote{More generally, $\phi_8/\mu_{\E\E}$ could differ from 1, and the relations
(\ref{eq:BM_parameters}) would be modified by factors of $\phi_8/\mu_{\E\E}$.  We allow for this more general form in the following.}\ \ 
The resulting contributions to the different gauge boson masses are shown in the third column of Table~\ref{tab:gauge_boson_masses}.  
Purely diagonal $\langle\Phi_8\rangle$ VEVs leave the $A_3$ and $A_8$ gauge bosons massless.  
To generate mass for $A_{8}$,  we introduce the VEV
\be \label{eq:Phi3_VEV}
    \langle\Phi_3\rangle = \phi_3\,(0,0,1)^T ,
\ee
which  further lifts the $A_{4,5,6,7}$ masses,  leaving only $A_3$ massless.
The numerical values of the contributions to $M_i^2/g'^2$ are shown in the fourth column of Table~\ref{tab:gauge_boson_masses}.
As we will see in the next section, phenomenological constraints on the leptophilic gauge bosons lead to the requirements $\phi_8\gtrsim 1$~TeV and $\phi_3\gtrsim 10$~TeV.

The $\Phi_6$ VEV contributes to all of the masses and causes mixing between the gauge bosons.
We assume that it is much smaller than the other two VEVs so that the mixing, as well as the shifts to the heavy masses, can be neglected, or computed perturbatively.
Therefore the $\langle\Phi_6\rangle_{3,3}$ 
element directly determines $M_3$, and thereby  the new physics contribution to $(g-2)_\mu$.

\begin{table}[t]\centering
\setlength\tabcolsep{5pt}
\def\arraystretch{1.5}
\begin{tabular}{|c|c|c|c|c|}\hline
\begin{tabular}[c]{@{}c@{}}Gauge\\boson\end{tabular} & \begin{tabular}[c]{@{}c@{}}Flavor\\structure\end{tabular} & $\Phi_8$ & $\Phi_3$ & $\Phi_6$\\
\hline
$A_{1,2}$ &$\phantom{1\over\sqrt{3}}$\scalebox{0.6}{ $\begin{pmatrix} 0 & * & 0\\ * & \phantom{-}0\phantom{-} & 0\\ 0 & 0 & 0\end{pmatrix}\begin{array}{c}
\tau\\ \mu\\ e\end{array}$}
    & $4\,\phi_8^2$ & \xmark & \cmark \\
\hline
$A_{3}$ & $\phantom{1\over\sqrt{3}}$
\scalebox{0.6}{$ \begin{pmatrix} 1 & 0 & 0\\ 0 & -1\phantom{-} & 0\\ 0 & 0 & 0\end{pmatrix}\quad\;\;\;$}
& \xmark & \xmark  & \cmark\\
\hline
$A_{4,5}$ & $\phantom{1\over\sqrt{3}}$\scalebox{0.6}{$\begin{pmatrix} 0 & 0 & *\\ 0 & \phantom{-}0\phantom{-} & 0\\ * & 0 & 0 \end{pmatrix}$\quad\;} & $\phi_8^2$ & $\sfrac12\, \phi_3^2$ & \cmark\\
\hline
$A_{6,7}$ & $\phantom{1\over\sqrt{3}}$\scalebox{0.6}{$\begin{pmatrix} 0 & 0 & 0\\ 0 & \phantom{-}0\phantom{-} & *\\ 0 & * & 0 \end{pmatrix}$\quad\;} & $\phi_8^2$ & $\sfrac12\, \phi_3^2$ &\cmark\\
\hline
$A_{8}$ & ${1\over\sqrt{3}}$\scalebox{0.6}{ $\begin{pmatrix} 1 & \;\;0\; & 0\\ 0 & \;\;1\; & 0\\ 0 & \;\;0\; & -2\end{pmatrix}\quad\;\;$} & \xmark & $\sfrac23\, \phi_3^2$ & \cmark\\
\hline
\end{tabular}
\caption{Flavor structure of the couplings of the 8 leptophilic gauge bosons in the SU(3)$_\ell$ model, together with the contributions to $M_{i}^2/g'^2$, their squared masses divided by $g^2$, arising from the various scalar VEVs (see Eq.~\ref{eq:Phi8_VEV}).
The contributions from  $\Phi_6$ are small compared to those of $\Phi_3$ and $\Phi_8$ by construction, and they lead to mixing between the tabulated Lagrangian states.}
\label{tab:gauge_boson_masses}
\end{table}

Using Eq.~(\ref{eq:sextet_VEV}), the light gauge boson mass can be written as
\be
    M_3 \sim g' {m_D^2\over y_\N\, m_{\nu_1}} \equiv
        g' \phi_6\,,
\ee
where the exact proportionality factor depends on the absolute scale of neutrino masses and varies between $\sim 0.5$ and $1.5$ within the range of interest.
It will be shown in the next section that for values of $g'$ and $M_3$ that can explain the muon anomalous magnetic moment,  $\phi_6 = \mathcal{O}(20-200)\,$GeV. 
If $y_\N\sim 1$, this is also the scale of the sterile neutrino masses,  whose mass matrix is $m_\N = y_\N\langle\Phi_6\rangle$.
In the following sections, we discuss the phenomenological consequences of the new particles predicted by the model.

%%%%%%%%%%%%%%%%%%%%%%%%%%%%%%%%%%%%%%%%%%%%%%%%%%%%%%%%%%%%%%%%%%%%%%
%%%%%%%%%%%%%%%%%%%%%%%%%%%%%%%%%%%%%%%%%%%%%%%%%%%%%%%%%%%%%%%%%%%%%%

\section{Phenomenology of the leptophilic gauge bosons}
\label{sect:pheno}
The phenomenology of gauge bosons associated with lepton number gauge symmetries has been widely studied; see e.g.~\cite{Foldenauer:2016rpi,Buras:2021btx} and references therein.
In the SU(3)$_\ell$ model presented here, neglecting the small mixing between gauge bosons,  there are six purely flavor-violating currents, mediated by $A_{1,2}$ for $\mu\leftrightarrow\tau$, $A_{4,5}$ for $e\leftrightarrow\tau$, and $A_{6,7}$ for $e\leftrightarrow\mu$.
The flavor-conserving $A_3$ couples to $\mu-\tau$, and $A_8$ to the $2e-\mu-\tau$ current.
These leptophilic gauge bosons have implications for a number of different observables.

%%%%%%%%%%%%%%%%%%%%%%%%%%%%%%%%%%%%%%%%%%%%%%%%%%%%%%%%%%%%%%%%%%%%%%

\subsection{Muon anomalous magnetic moment}
\label{amusect}

\begin{figure}[t]
\centering
\includegraphics[width=0.47\textwidth]{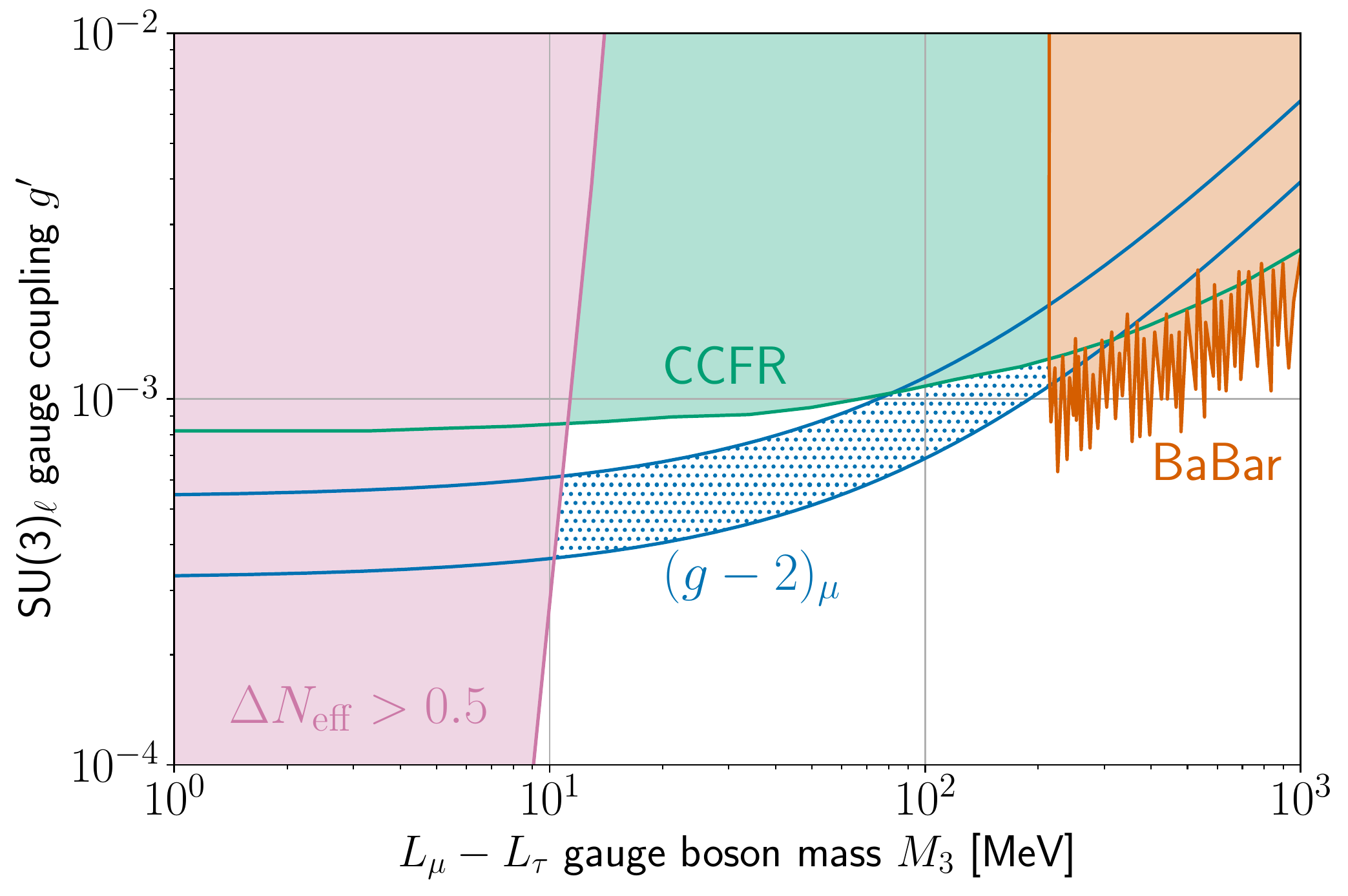}
\caption{Region of the gauge coupling versus mass parameter space for which a massive dark photon coupled to the $L_\mu-L_\tau$ current can explain the observed value of the muon $g-2$ factor.
The blue-dotted region shows the $2\sigma$ preferred values obtained from the recently reported measurement~\cite{Muong-2:2021vma} and SM calculations~\cite{Aoyama:2020ynm}. The other shaded areas correspond to constraints from Babar~\cite{BaBar:2016sci} (orange), CCFR~\cite{Altmannshofer:2014pba} (green), and cosmology~\cite{Escudero:2019gzq} (purple).}
 \label{fig:gminus2}
\end{figure}

By construction, $A_3$ is much lighter than the rest of the vector bosons, due to the symmetry-breaking pattern of the model.
This enables it to explain the current $4.2\sigma$ discrepancy between experimental measurements~\cite{Muong-2:2006rrc,Muong-2:2021vma} and SM predictions~\cite{Aoyama:2020ynm} of the anomalous magnetic moment of the muon.\footnote{A recent lattice evaluation of $(g-2)_{\rm SM}$~\cite{Borsanyi:2020mff} may help alleviate this tension, but it is in disagreement with the other SM predictions~\cite{Aoyama:2020ynm}.}\ \
A gauge boson $A'$ coupling to the muonic vector current with strength $g'$ contributes to $a_\mu = (g-2)_\mu/2$ at the one-loop level~\cite{Baek:2001kca} with
\begin{equation}
\Delta a_\mu = \frac{g'^2}{4\pi^2}\int_0^1 \mathop{\diff x} \frac{m_\mu^2\, x\, (1-x)^2}{m_\mu^2 (1-x)^2 + m_{A'}^2 \,x}.
\end{equation}
This has the right sign to address the experimental anomaly and is phenomenologically viable as long as $A'$ couples negligibly to electrons, which is the case for $A_3$.
Such a $L_\mu-L_\tau$ gauge boson was searched for in the $e^+e^- \rightarrow 4\mu$ channel at BaBar~\cite{BaBar:2016sci} and through neutrino trident production at CCFR~\cite{Altmannshofer:2014pba}.
Taking the ensuing constraints into account, the $A_3$ gauge boson can explain the $(g-2)_\mu$ discrepancy as long as its mass lies in the $10\lesssim m_{A_3}\lesssim 200$~MeV range and the gauge coupling is $3\times 10^{-4}\lesssim g' \lesssim 10^{-3}$ \cite{Holst:2021lzm,Hapitas:2021ilr} as
shown in Fig.~\ref{fig:gminus2}.

%%%%%%%%%%%%%%%%%%%%%%%%%%%%%%%%%%%%%%%%%%%%%%%%%%%%%%%%%%%%%%%%%%%%%%

\begin{figure}[t]
\centering
\includegraphics[width=0.47\textwidth]{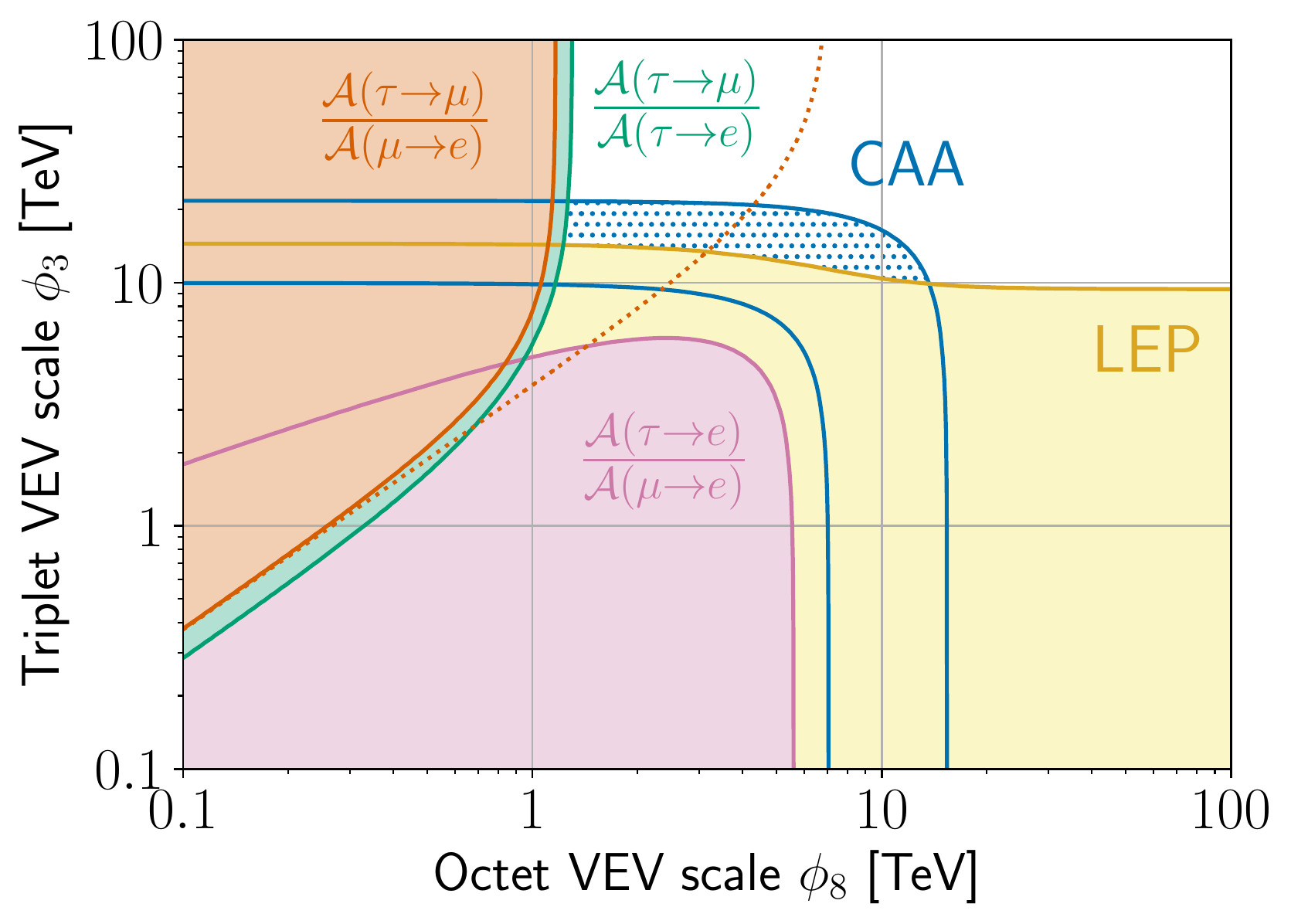}
\caption{$95\%$ CL contours on the model parameters $\phi_8$ and $\phi_3$ from LEP dilepton analyses and lepton flavor universality constraints as labelled in the figure. The blue-dotted region shows the $2\,\sigma$ preferred values to address the CKM unitarity deficit, also known as the Cabibbo Angle Anomaly (CAA). The dotted orange line shows the edge of the preferred region for the slight $\sim 2\,\sigma$ excess in the LFU ratio~\eqref{eq:luv_2}.}
 \label{fig:lepton_universality}
\end{figure}

\subsection{Di-lepton searches at LEP}

The LEP experiments searched for $e^+e^-\to \ell^+\ell^-$
events beyond the Standard Model \cite{ALEPH:2013dgf}, parametrizing the new physics (NP) contributions by contact interactions of the form
\be
    {4\pi\over\Lambda^2} J_{e,\alpha}
    \left[\sfrac12 J_e^\alpha + J_\mu^\alpha + J_\tau^\alpha
    \right],
\ee
where $J_i^\mu$ is the vector current for charged leptons of flavor $i$. 
Different limits on $\Lambda$ were derived depending on the combinations of final states observed.  
However, the derived limits do not directly apply to states such as $A_8$ that couple with different strengths to different flavors. 

However, Ref.\ \cite{ALEPH:2013dgf}  provides the observed differential distributions for the $e^+e^-\rightarrow \ell\ell$ processes studied by LEP, enabling us to
derive limits on nonabelian leptophilic gauge bosons
such as those in the present model.
For dimuon and ditau final states we use the data for the total cross-section $\sigma_{\rm T}$ and forward-backward asymmetry $A_{\rm fb}$ as a function of $\sqrt{s}$, while for the $e^+e^-$ final state we use the full averaged differential cross section given in~\cite{ALEPH:2013dgf}.
Employing a $\chi^2$ test statistic on the SM versus SM + NP models leads to the following bounds on the individual gauge boson masses at $95\%$ CL:
\bea
    {M_{4,5}\over g'} &>& 4.6\,{\rm TeV},\nn\\
    {M_{6,7}\over g'} &>& 5.2\, {\rm TeV},\nn\\
    {M_8\over g'} &>& 8.9\,{\rm TeV}.
    \label{LEPbounds2}
\eea

We further construct the combined $\chi^2$ for all the aforementioned channels, including the effects of all relevant gauge bosons parametrized by the two model parameters $\phi_3$ and $\phi_8$.
Doing so leads to the $95\%$ CL constraints on the VEVs corresponding to the yellow contours in Fig.~\ref{fig:lepton_universality}.
As shown in Eq.~(\ref{LEPbounds2}), the strongest limits are on $A_8$, it being the only state coupling diagonally to the electron current.
Since the mass of $A_8$ comes solely from $\langle\Phi_3\rangle$, this constrains $\phi_3$ independently of the value of $\phi_8$, which explains the shape of the exclusion contour in Fig.~\ref{fig:lepton_universality}.

%%%%%%%%%%%%%%%%%%%%%%%%%%%%%%%%%%%%%%%%%%%%%%%%%%%%%%%%%%%%%%%%%%%%%%

\subsection{Lepton flavor universality violation}

Vector bosons with flavor off-diagonal couplings can induce lepton decays of the form $\ell_i\rightarrow\ell_j \bar{\nu}\nu$, which interfere with the SM amplitudes and spoil the flavor-universal nature of these weak decays.
Following the prescription of~\cite{Buras:2021btx}, these effects can be measured using the amplitude ratios
\begin{equation}
    R(\ell_i\rightarrow\ell_j) = \frac{\mathcal{A}(\ell_i\rightarrow\ell_j \bar{\nu}\nu)}{\mathcal{A}(\ell_i\rightarrow\ell_j \bar{\nu}\nu)_{\rm SM}} = 1 + 2 \frac{g'^2}{g_2^2} \frac{m_W^2}{m_{A'}^2}\,,
\end{equation}
where $g_2$ denotes the SU(2)$_L$ gauge coupling and we have expanded to first order in $g'^2$, {\it i.e.,} keeping only SM-NP interference terms.
These can be confronted with experimental measurements of the ratios of partial widths in purely leptonic decays~\cite{HFLAV:2019otj},\footnote{The $\tau-\mu$ universality ratio can also be measured in semi-hadronic decays, but with a larger uncertainty that does not significantly strengthen the constraints.}
\begin{align}\label{eq:luv_1}
    \frac{\mathcal{A}(\tau\rightarrow\mu \bar{\nu}\nu)}{\mathcal{A}(\tau\rightarrow e \bar{\nu}\nu)} &= 1.0018\pm 0.0014 , \\
    \label{eq:luv_2}
    \frac{\mathcal{A}(\tau\rightarrow\mu \bar{\nu}\nu)}{\mathcal{A}(\mu\rightarrow e \bar{\nu}\nu)} &= 1.0029\pm 0.0014 , \\
    \label{eq:luv_3}
    \frac{\mathcal{A}(\tau\rightarrow e \bar{\nu}\nu)}{\mathcal{A}(\mu\rightarrow e \bar{\nu}\nu)} &= 1.0010\pm 0.0014 ,
\end{align}
with the correlations given in Ref.~\cite{HFLAV:2019otj}.
In our model, the $R$-ratios are completely determined by the $\Phi_3$ and $\Phi_8$ VEVs, which can be parametrized using $\phi_3$ and $\phi_8$ as defined in Eqs.~\eqref{eq:Phi8_VEV} and~\eqref{eq:Phi3_VEV}.
Applying the experimental constraints in Eqs.~\eqref{eq:luv_1}-\eqref{eq:luv_3} at $95\%$ CL leads to the exclusion limits shown in Fig.~\ref{fig:lepton_universality}.

\subsection{Cabibbo angle (or CKM) anomaly}

In addition to LFU violation, a modification of the $\mu\rightarrow e \bar{\nu}\nu$ decay rate changes the inferred value of the Fermi constant, which affects the determination of $V_{ud}$ from $\beta$ decays.
The latter measurements currently exhibit a $\sim 3 \sigma$~\cite{Belfatto:2019swo,Grossman:2019bzp,Shiells:2020fqp,Seng:2020wjq} tension with the SM predictions, which could be eased by a NP contribution at the level of~\cite{Buras:2021btx}
\begin{equation}
R(\mu\rightarrow e) = 1.00075\pm 0.00025\,.
\end{equation}
This experimental anomaly was originally formulated as a breakdown of unitarity in the first row of the CKM matrix, and is thus commonly referred to as the Cabibbo angle anomaly (CAA).
Since the sum $|V_{ui}|^2$ appears to be less than unity, one interpretation is that weak decays of nuclei are suppressed relative to those of leptons.
The additional contributions of $A_{6,7}$ gauge interactions to muon decays can explain the anomaly if the effective Fermi constant $G_\mu$ inferred from muon decays is increased by a factor of $(1+\delta_\mu)$, with $\delta_\mu = 7\times 10^{-4}$~\cite{Belfatto:2019swo}.
It can be achieved in the present model with
\be
    {M_{6,7}\over g'}\cong 13\,{\rm TeV}\,
    \label{CAAeq}
\ee
due to the additional neutral current contribution to $\mu\to e\nu_\mu\bar\nu_e$.
This is consistent with the LEP and LFU bounds, as can be seen in Fig.~\ref{fig:lepton_universality}, where the $2\sigma$ preferred region to explain the CAA anomaly corresponds to the blue-dotted contour.
If a NP origin of the anomaly were to be confirmed, it could be explained by the nonabelian model with VEVs $\phi_3\sim10-20$~TeV and $\phi_8\sim1-15$~TeV.\footnote{The LFU ratio~\eqref{eq:luv_2} also deviates by $2\sigma$ from the SM prediction, which would single out an even more restricted range $\phi_8\sim1-5$~TeV for the $\Phi_8$ VEV as indicated by the orange dotted line in Fig.~\ref{fig:lepton_universality}.}

%%%%%%%%%%%%%%%%%%%%%%%%%%%%%%%%%%%%%%%%%%%%%%%%%%%%%%%%%%%%%%%%%%%%%%

\subsection{Gauge boson mixing}
The $\Phi_6$ VEV generates mixing between all the gauge boson Lagrangian states in the model, with mixing angles of order $\sin\xi_{ij}\sim(\phi_6/\phi_{3,8})^2\lesssim 10^{-4}$ for the VEVs of interest here.
To see that these are phenomenologically harmless, we consider the most sensitive observable probing transitions enabled by them, $\mu\rightarrow e \gamma$, whose branching ratio is constrained to be $\mathrm{Br}(\mu\rightarrow e\gamma)\leq 4.2\times 10^{-13}$ \cite{SINDRUMII:2006dvw,BaBar:2009hkt,MEG:2016leq}.
This process is generated at one loop via $A_{1,2}-A_{4,5}$ mixing (internal $\tau$ lepton) or $A_{6,7}-A_{8}$ mixing (internal electron and muon).
Following Ref.\ \cite{Buras:2021btx}, we derive the bounds
\begin{align}
\frac{M_{i,j}}{g'} &\geq 3.7\,\mathrm{TeV}\,\sqrt{ \frac{\sin\xi_{ij}}{10^{-4}} }\,,\nn\\
\frac{M_{k,8}}{g'} &\geq 48\,\mathrm{GeV}\,\sqrt{ \frac{\sin\xi_{k8}}{10^{-4}} }\,,
\end{align}
where $i=1,2$, $j=4,5$, and $k=6,8$.
The first limit is stronger than the second one by a factor of $\sqrt{m_\tau/m_e}\sim 60$ due to a cancellation of the $\mu$-exchange contribution in the latter.
In any case, neither of the two compete with the previously discussed LEP or LFU bounds.

%%%%%%%%%%%%%%%%%%%%%%%%%%%%%%%%%%%%%%%%%%%%%%%%%%%%%%%%%%%%%%%%%%%%%%
%%%%%%%%%%%%%%%%%%%%%%%%%%%%%%%%%%%%%%%%%%%%%%%%%%%%%%%%%%%%%%%%%%%%%%

\section{Phenomenology of the heavy neutral leptons}\label{sec:HNL}

\begin{figure}[t]
\centering
\includegraphics[width=0.49\textwidth]{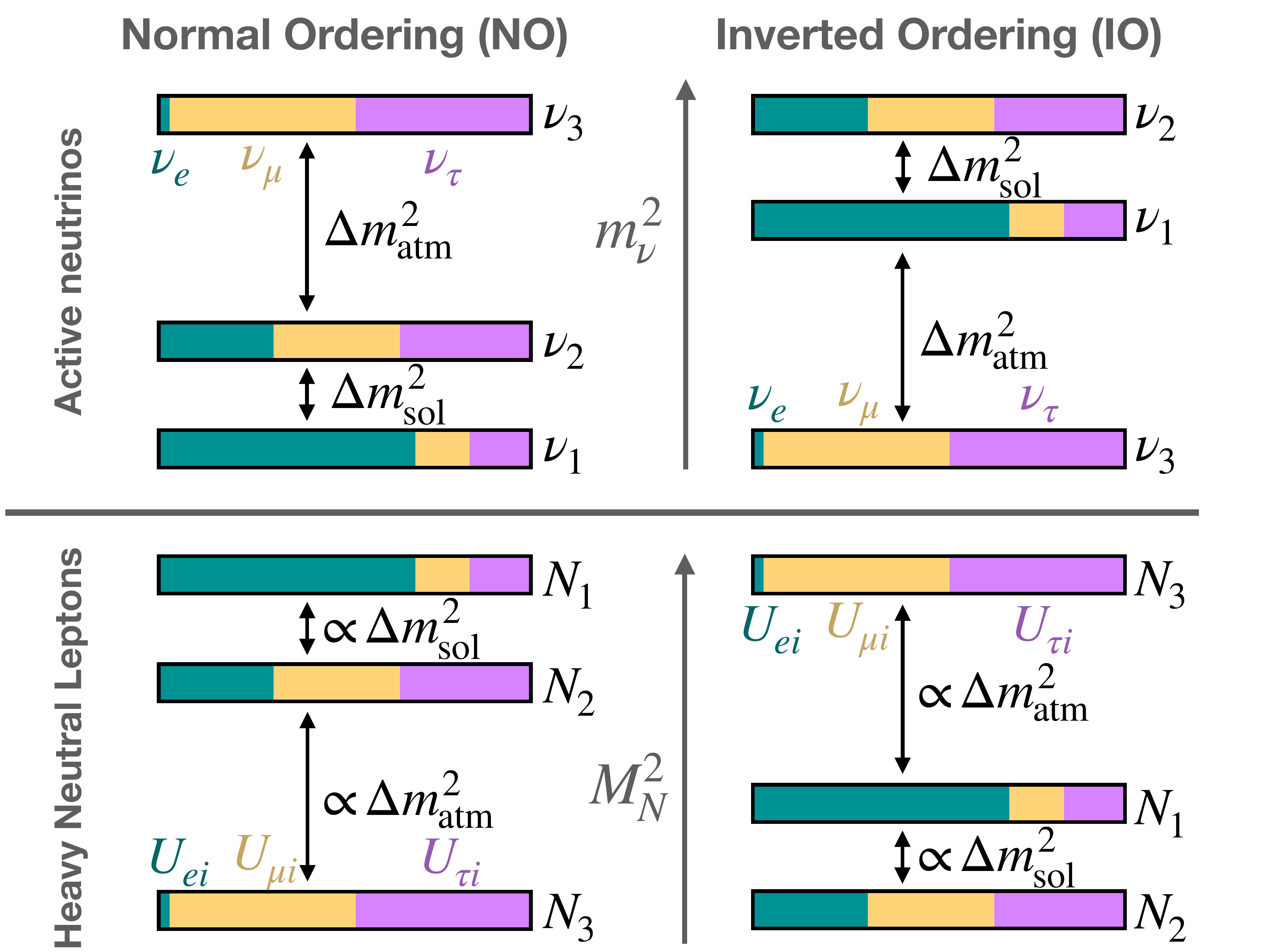}
\caption{Schematic representation of the mass and mixing hierarchies of the active neutrinos and the heavy neutral leptons in the gauged SU(3)$_\ell$ model, for both normal and inverted orderings.
The masses of the HNLs are predicted to be inversely proportional to the light neutrino ones, so that they are distributed following the reflected pattern represented in the figure.
The mixings of each HNL mass eigenstate with the light neutrino flavors, represented by colors, exactly match the flavor admixture of the corresponding active neutrino, which is dictated by the PMNS matrix.}
 \label{fig:mass_hierarchy}
\end{figure}

The spectrum of heavy neutral leptons $N_i$, and their mixing with the light neutrinos $\nu_i$, is largely dictated by the measured $\nu_i$ masses and mixings, once the lightest mass $m_{\nu, l}$ is specified. 
In particular, there is a strict relation between pairs of $m_{\nu_i}$ and $M_{N_i}$ eigenvalues,
\be
    m_{\nu_i}M_{N_i} = m_D^2, \quad {i=1,2,3}\,,
    \label{mMrel}
\ee
where $m_D$ is the universal neutrino Dirac mass.
Because of SU(3)$_\ell$ symmetry, $m_D$ is just a number, not a matrix.
This implies that the logarithmic range of $N_i$ masses is the same as that of the light neutrinos, as is represented schematically in Fig.~\ref{fig:mass_hierarchy}.
The mixing angles between $N_i$ and the $\nu_i$ are diagonal in the basis of mass eigenstates, 
and are approximately
\be
    U_i = {m_D\over M_{N_i}} = \sqrt{\frac{m_{\nu_i}}{M_{N_i}}}\,.
\ee
Thus, heavier HNLs are more weakly mixed and the flavor mixing pattern of each one exactly matches that of the corresponding active neutrino, as is illustrated in Fig.~\ref{fig:mass_hierarchy}.
We now proceed to quantify this picture more precisely.

%%%%%%%%%%%%%%%%%%%%%%%%%%%%%%%%%%%%%%%%%%%%%%%%%%%%%%%%%%%%%%%%%%%%%%

\subsection{HNL masses}
There are three parameters that determine the HNL mass and mixing spectrum, together with the $\Phi_6$ contribution to gauge boson masses: the Yukawa couplings $y_N$ and $y_{\L\N}$ and the mass of the lightest active neutrino $m_{\nu,l}$.
This state is labeled as $\nu_1$ in the normal
mass hierarchy (NO), and $\nu_3$ in the inverted one (IO),
as shown in Fig.\ \ref{fig:mass_hierarchy}.
The choice of hierarchy as well as the relative signs of active neutrino masses affects the HNL mixings.
 
\begin{figure}[t]
\centering
\centerline{
\includegraphics[width=0.49\textwidth]{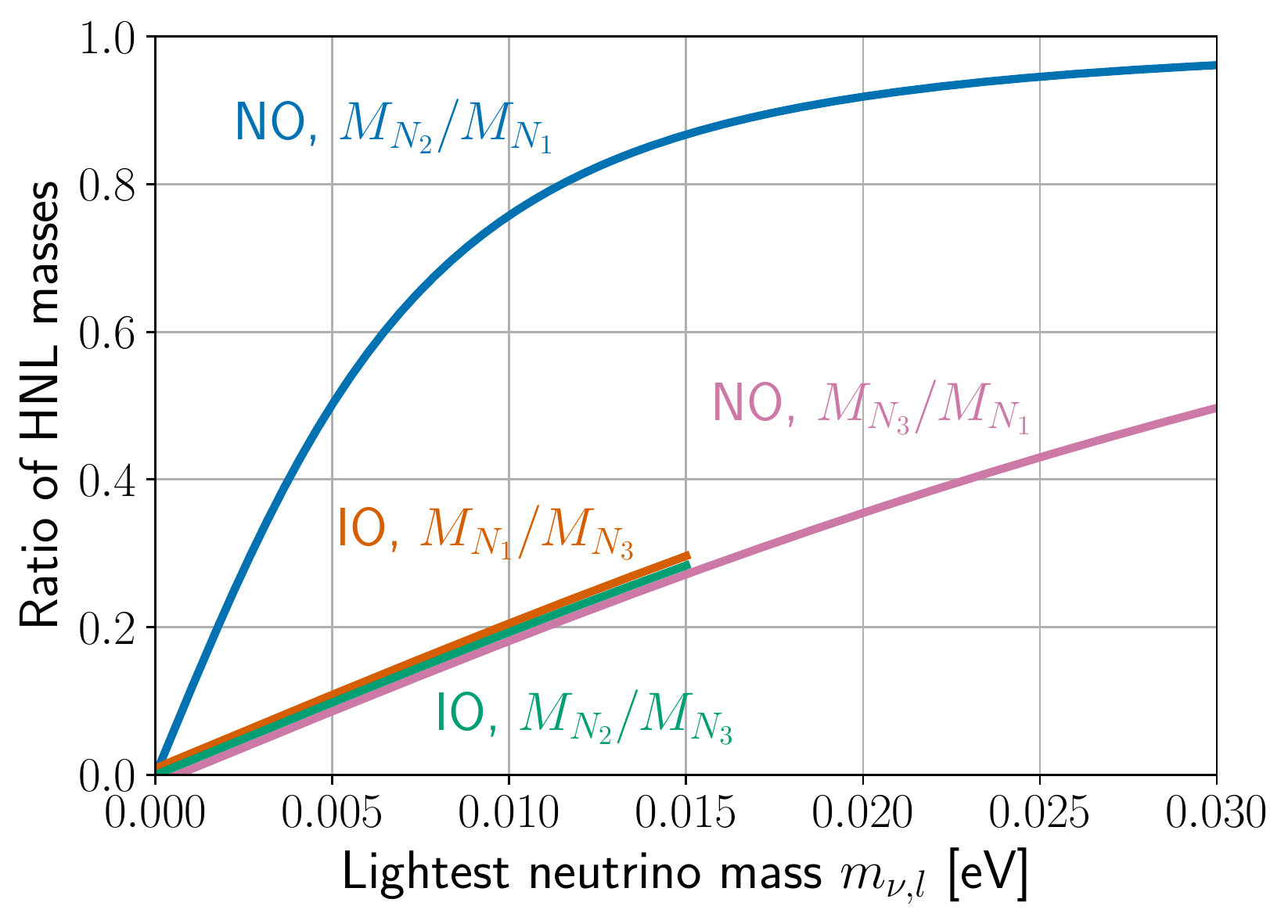}}
\caption{Ratios of the absolute value of the lighter two HNL masses to the heaviest one, as a function of the mass of the lightest active neutrino.
For normal ordering, there is a single HNL that is substantially lighter than the other two, while for inverted ordering (IO) there are two nearly degenerate HNLs that are lighter than the third one.  For IO, the curves terminate when the cosmological limit on $\sum_i |m_{\nu,i}|$ becomes saturated. }
 \label{fig:fs}
\end{figure}

From Eq.~\eqref{eq:sextet_VEV}, the sextet VEV can be written as
\begin{equation}
    \langle \Phi_6 \rangle = - \frac{y_{\L\N}^2}{y_N} \, \frac{\bar{v}^2}{m_{\nu,l}} \, U_{\rm\sss PMNS}\, \hat{D}^{-1}\, U_{\rm\sss PMNS}^T\,,
\end{equation}
where $\hat{D}$ is the diagonal active neutrino mass matrix rescaled by $1/m_{\nu,l}$ and $U_{\rm\sss PMNS}$ is the PMNS matrix with columns arranged according to our flavor ordering scheme $(\tau,\,\mu,\,e)$.
The $\Phi_6$ VEV determines the ratio $M_{3}/g'$ for the $\mu-\tau$ gauge boson via Eq.~\eqref{gbmasses}, and thus its ability to explain the $(g-2)_\mu$ anomaly.
As was noted in the previous section,
\begin{equation}
    \frac{M_3}{g'}\in (18\,\mathrm{GeV},\,200\,\mathrm{GeV})
\end{equation}
is the preferred range at the $2\sigma$ level, corresponding to $g'\in(3.7,\,13)\times10^{-3}$.
The precise value of $M_3$ predicted in this way has a weak dependence on the relative sign of the active neutrino masses, which amounts to a factor of $\sim 2$ for $m_{\nu,l}=0.03$~eV and disappears for  $m_{\nu,l}\ll 0.03$\,eV.

\begin{figure*}[t]
\centering
\centerline{
\includegraphics[width=0.49\textwidth]{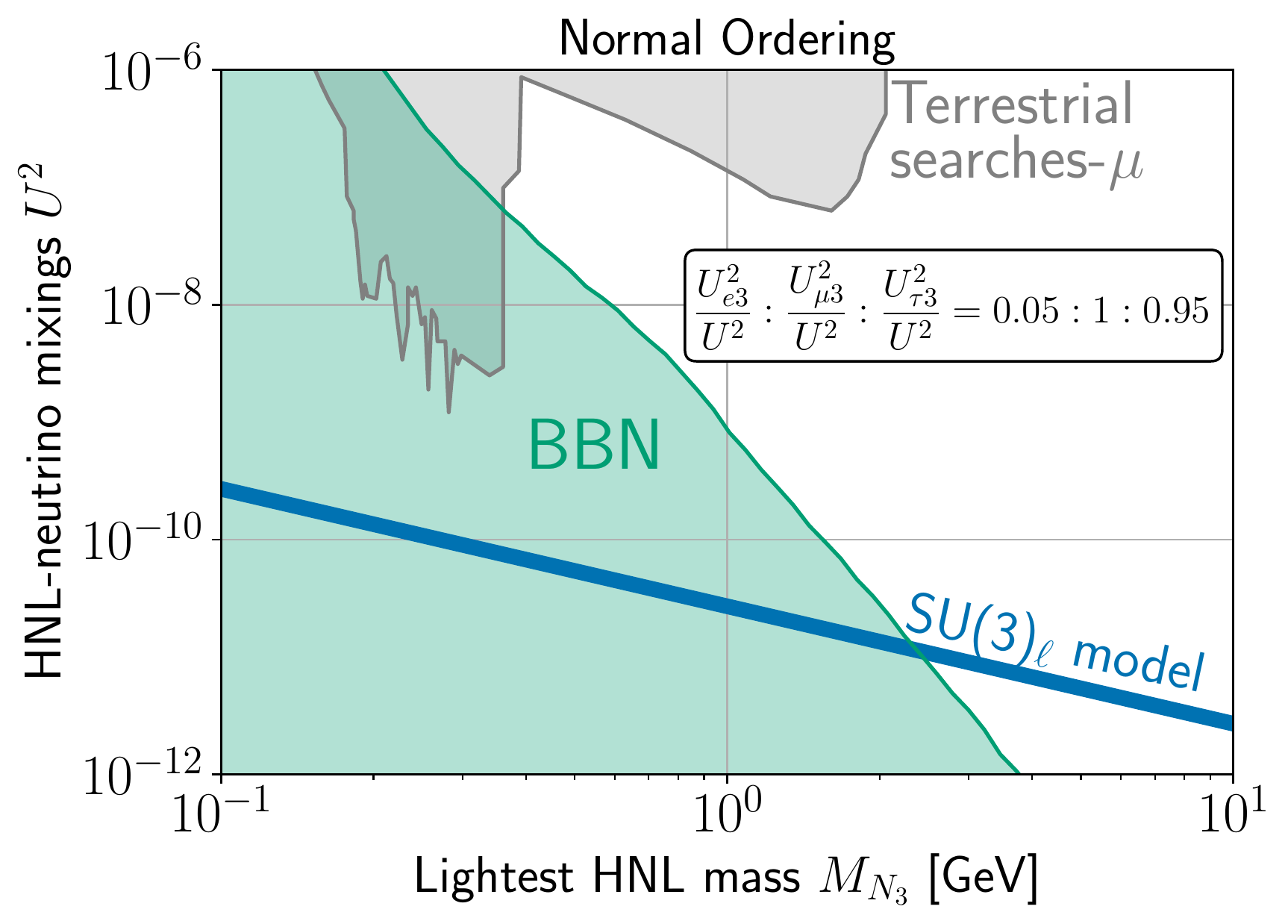}\hfil
\includegraphics[width=0.49\textwidth]{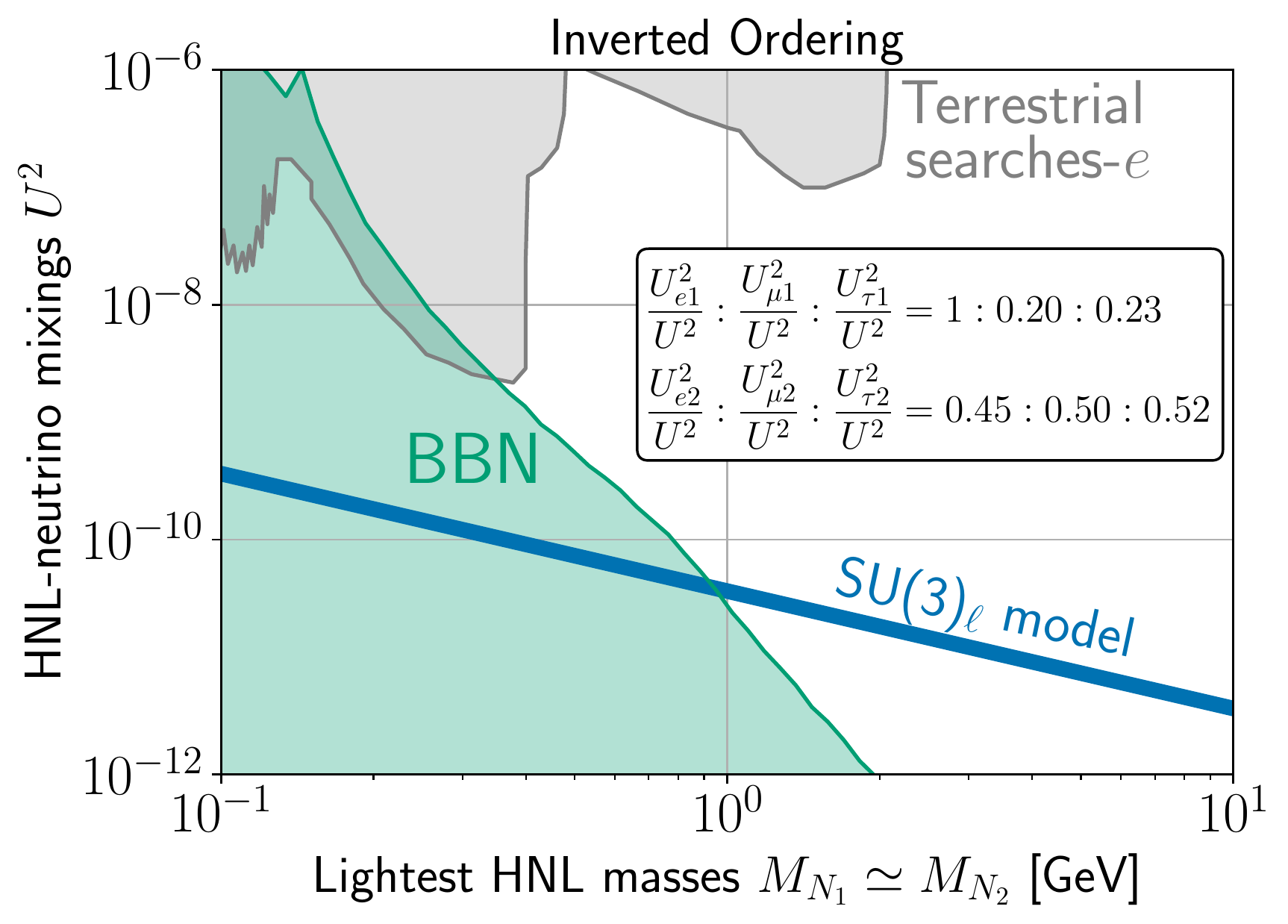}}
\caption{Dark blue: predictions from the gauged SU(3)$_\ell$ lepton flavor model for the mass and mixing of light HNLs.
The mixings are normalized to the muon \emph{(left)} or electron \emph{(right)} one as indicated in the insets.
We show bounds (grey) from terrestrial searches as compiled in~\cite{Alekhin:2015byh} as well as the BBN limit (green) computed following~\cite{Boyarsky:2020dzc,Bondarenko:2021cpc}.
For the normal  mass hierarchy \emph{(left)}, there is a single light HNL with suppressed $e$ mixing and we therefore show terrestrial searches based on the $\mu$ mixing. For inverted hierarchy \emph{(right)}, there are two quasi-degenerate light HNLs with fairly universal mixings and the leading terrestrial constraints therefore come from $e$ mixing.}
 \label{fig:HNL_mixing_BBN}
\end{figure*}

The mass matrix for the HNLs is in turn determined by the sextet VEV as
\bea\label{eq:HNL_mass}
    M_N &=& y_N\langle\Phi_6\rangle\nn\\
    &=& - y_{\L\N}^2  \frac{\bar{v}^2}{m_{\nu,l}} \, U_{\rm\sss PMNS}\, \hat{D}^{-1}\, U_{\rm\sss PMNS}^T\,.
\eea
Using the top line in Eq.~\eqref{eq:HNL_mass}, the $(g-2)_\mu$-preferred values for $\langle\Phi_6\rangle$ result in the prediction that the scale of HNL masses lies in the $\mathcal{O}(1-100) \mathrm{GeV}$ range for $\mathcal{O}(1)$ values of $y_N$.

Eq.~\eqref{eq:HNL_mass} implies that the PMNS matrix also represents the unitary transformation that diagonalizes $M_N$.
As a consequence, each HNL mass eigenstate is strictly inversely proportional to the corresponding active neutrino mass,
\begin{equation}
    M_{N_i} = - y_{\L\N}^2  \frac{\bar{v}^2}{m_{\nu_i}}\,.
\end{equation}
This means that in the normal ordering case, there is one light and two heavy HNLs, while for inverted ordering there are two light and one heavy HNLs.
This is schematically represented in Fig.~\ref{fig:mass_hierarchy}, while the quantitative mass ratios are shown in Fig~\ref{fig:fs} as a function of $m_{\nu,l}$.
The largest considered value of $m_{\nu, l}=0.03\, (0.015)$~eV in the NO (IO) case, roughly saturates the cosmological constraint on the sum of the absolute value of neutrino masses\footnote{A recent evaluation~\cite{DiValentino:2021hoh} finds a tighter limit $\sum |m_{\nu_i}|\leq0.09$~eV, which would compromise the viability of the IO scenario but would not significantly affect the NO one.} $\sum |m_{\nu_i}|\leq0.12$~eV~\cite{Planck:2018vyg}.

%%%%%%%%%%%%%%%%%%%%%%%%%%%%%%%%%%%%%%%%%%%%%%%%%%%%%%%%%%%%%%%%%%%%%%

\subsection{HNL mixings}

\begin{figure*}[t]
\centering
\centerline{
\includegraphics[width=0.49\textwidth]{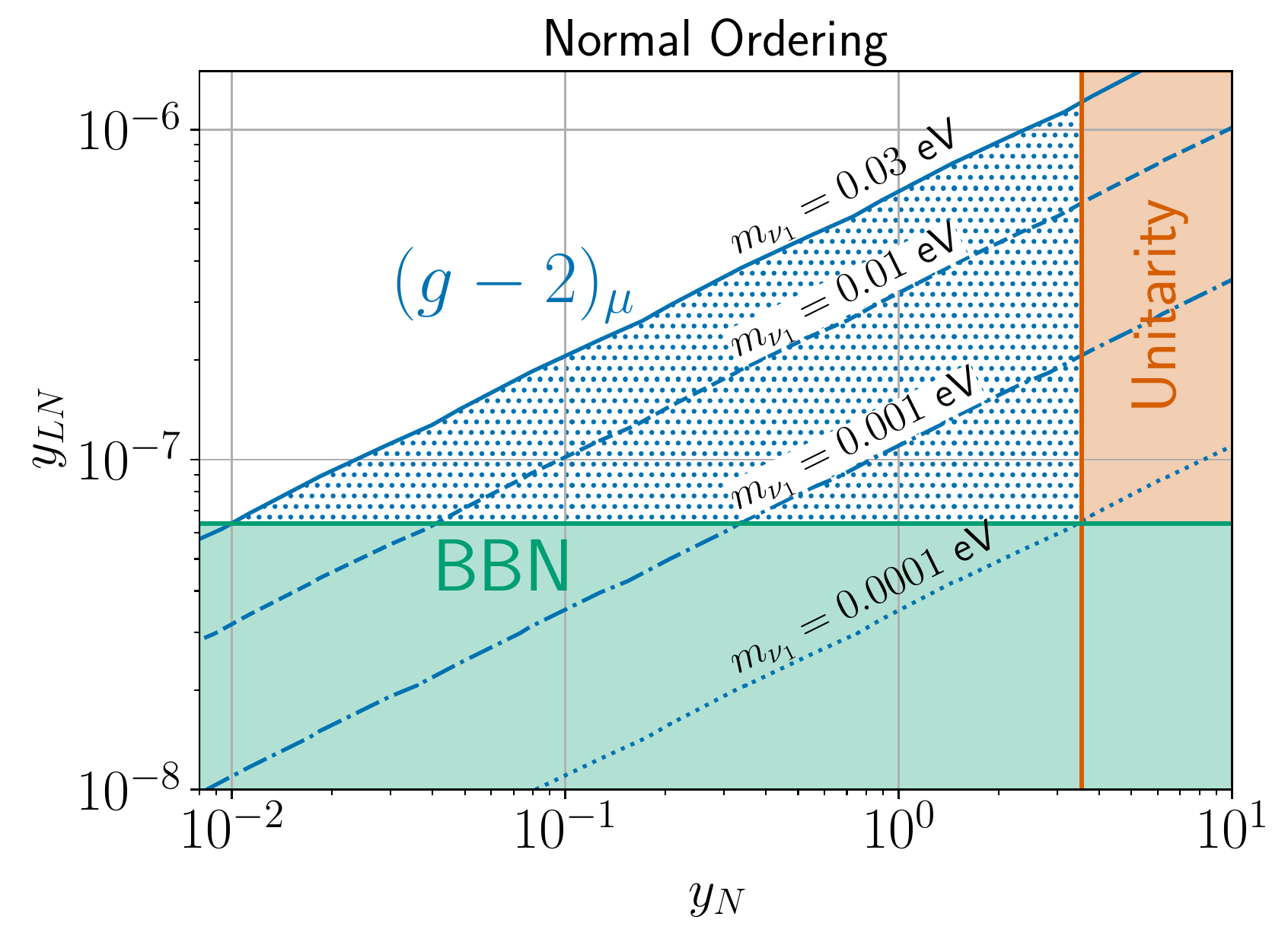}\hfil
\includegraphics[width=0.49\textwidth]{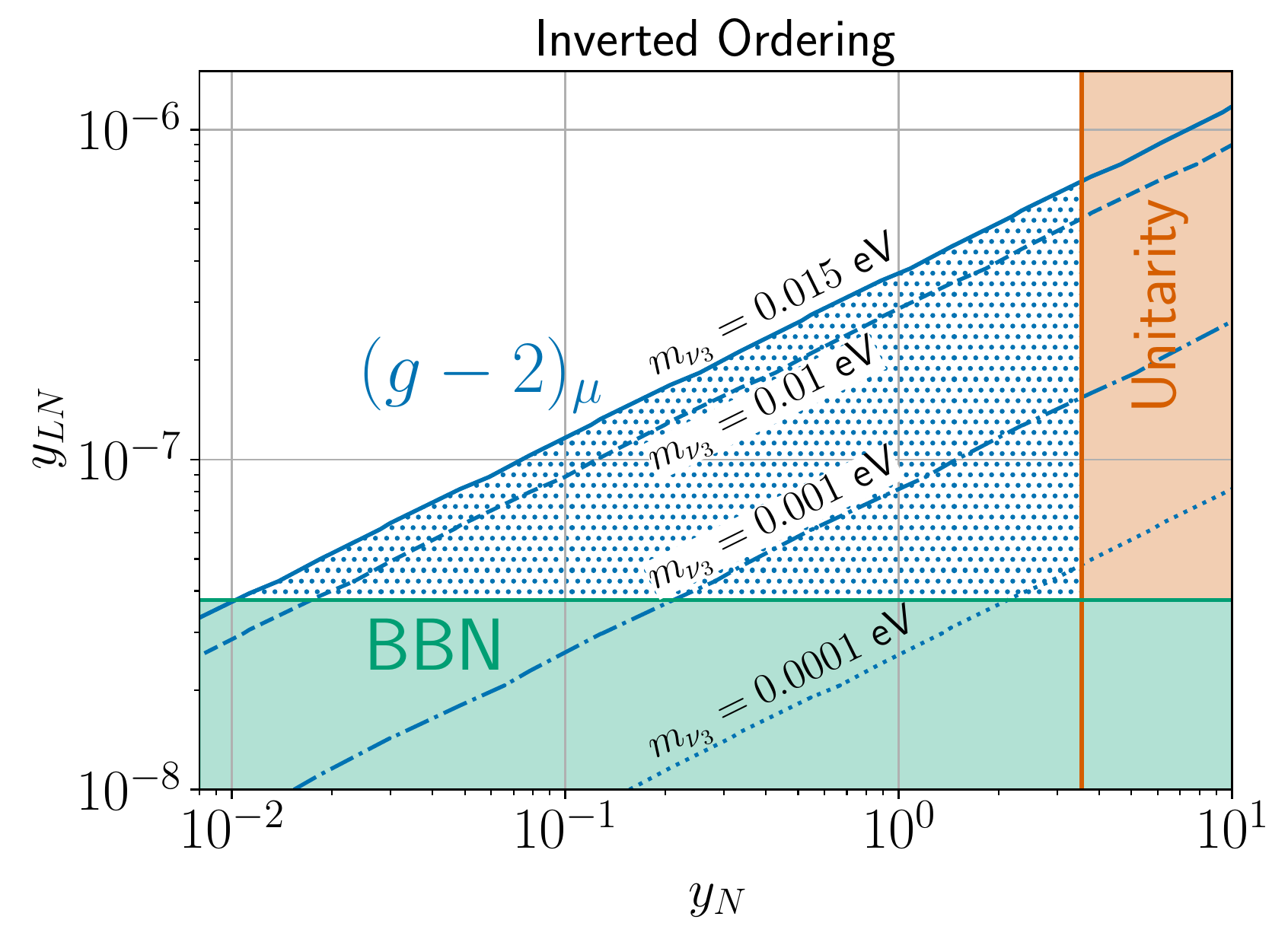}}
\caption{Regions of the $y_N$-$y_{\L\N}$ parameter space for which the gauged SU(3)$_\ell$ model can address the $(g-2)_\mu$ anomaly while satisfying all other phenomenological constraints, for various values of the lightest neutrino mass as labeled in the figure.
The blue-dotted regions show where the $A_3$ gauge boson mass and coupling $g'$ are consistent with the $2\sigma$ preferred values shown in Fig~\ref{fig:gminus2}.
The green region is the BBN exclusion from Fig.~\ref{fig:HNL_mixing_BBN} and the orange one
is forbidden by perturbative unitarity of the HNL Yukawa coupling, $y_N\leq\sqrt{4\pi}$.
Left (right) panel corresponds to the  NO (IO) active neutrino mass hierarchy, respectively.
}
\label{fig:HNL_yN_yLN}
\end{figure*}

Transforming to the HNL mass eigenbasis (denoted by primes), it is straightforward to show that the mixing between the light neutrino flavors and the HNLs is given by
\bea
    \nu_\alpha &\cong&\phantom{-} \nu'_\alpha + U_{\alpha i}\, N_i'\nn\\
    N_i &\cong& -U_{\alpha i}\,\nu'_\alpha + N'_i 
    \,,
\eea
where 
\bea
    U_{\alpha i} = U^{\mathrm{\sss PMNS}}_{\alpha i} \frac{m_{\nu_i}}{y_{\L\N}\bar{v}}\,.
\eea
The mixing structure is completely determined by the PMNS matrix: the mixing pattern of the HNL mass eigenstate $i$ is proportional to the flavor composition of the corresponding active neutrino mass eigenstate (see Fig.~\ref{fig:mass_hierarchy} for a graphical representation).
The approximate scale of the mixing is $U =  (m_\nu/M_N)^{1/2}\sim 10^{-6}-10^{-5}$ for HNLs in the $1-100$~GeV mass range.

While the HNL masses and mixings predicted by the
SU(3)$_\ell$ model are out of reach for present and upcoming terrestrial experiments, they have significant implications for cosmology.
In particular, HNLs decaying too close to the time of Big Bang Nucleosynthesis (BBN) can affect the predictions for light element yields~\cite{Dolgov:2000pj,Dolgov:2000jw}.
The strongest current constraints come from the hadronic decays of the HNLs, which inject light mesons into the plasma that can act to imbalance the proton-to-neutron ratio at BBN~\cite{Boyarsky:2020dzc}.

Following the prescription of~ Ref.\ \cite{Bondarenko:2021cpc} together with the width calculations of Ref.~\cite{Bondarenko:2018ptm}, we obtain the constraints shown in Fig.~\ref{fig:HNL_mixing_BBN} for normal and inverted mass hierarchies.
For NO, there is a single light HNL whose mixings match those of $\nu_3$ (see Fig.~\ref{fig:mass_hierarchy}).
As a consequence, its $e$ mixing is suppressed by a factor of $\sim 5$ compared to the $\mu$ and $\tau$ mixings.
Then the BBN bound gives $M_{N_3}\gtrsim2.4$~GeV or $U^2_{\mu3}\lesssim 1\times 10^{-11}$, which translates into $y_{\L\N}\gtrsim 6.4\times10^{-8}$.
IO has two light HNL with quasi-degenerate masses and mixings dictated by the flavor composition of $\nu_1$ and $\nu_2$ (see Fig.~\ref{fig:mass_hierarchy}).
The BBN bound arising from the combined effect of both HNLs leads to $M_{N_1}\sim M_{N_2}\gtrsim0.95$~GeV and $U^2_{e1}\lesssim 4\times 10^{-11}$, implying that $y_{\L\N}\gtrsim 4.0\times10^{-8}$.

The BBN constraint on $y_{\L\N}$, together with the bound $y_N\leq\sqrt{4\pi}$ arising from perturbative unitarity in $N\bar{N}\rightarrow N\bar{N}$ scattering, rule out 
regions of the $y_{\L\N}$ versus $y_\N$ parameter space, including some relevant for explaining the  $(g-2)_\mu$ anomaly.
They are shown in Fig.~\ref{fig:HNL_yN_yLN} for several
choices of the lightest active neutrino mass and for the two possible neutrino mass orderings.
Here we assume that $m_{\nu_3}$ has opposite sign to $m_{\nu_1}$ and $m_{\nu_2}$, which gives the model the greatest latitude for consistency.
Other choices of relative signs differ by factors  $\lesssim 2$ for the extent of $(g-2)_\mu$ preferred regions.

Several conclusions can be drawn from  Fig.~\ref{fig:HNL_yN_yLN}.
The minimum value for $m_{\nu,l}$ that is compatible with the experimental measurement of $(g-2)_\mu$ is $m_{\nu,l}\simeq 10^{-4}$~eV.
%The model thus precludes an arbitrarily small
%mass for the lightest neutrino.
Consequently, the maximum possible mass for the heaviest HNL is $M_{N_1}\simeq 1.2$~TeV for NO and $M_{N_3}\simeq 1$~TeV for IO.
At the other extreme, values of $y_{\L\N}$ as large as $10^{-6}$ are possible if the mass of the lightest neutrino saturates the cosmological upper bound.
This reveals a novel connection between the scale of neutrino masses and the anomalous magnetic moment of the muon within the SU(3)$_\ell$ paradigm.
It constitutes a unique prediction of the proposal, which could exclude it definitively in light of more precise experimental data.

%%%%%%%%%%%%%%%%%%%%%%%%%%%%%%%%%%%%%%%%%%%%%%%%%%%%%%%%%%%%%%%%%%%%%%
%%%%%%%%%%%%%%%%%%%%%%%%%%%%%%%%%%%%%%%%%%%%%%%%%%%%%%%%%%%%%%%%%%%%%%

\section{Conclusions}
\label{sect:conc}

In this work, we have proposed a framework for gauging the vectorial SU(3) family symmetry of lepton flavor in a minimal way.
The construction can consistently reproduce the observed patterns of charged lepton masses and neutrino masses and mixings.
The new particle content of the model, in addition to the eight leptophilic gauge bosons, consists of a right-handed neutrino triplet and a triplet of vectorlike isosinglet charged lepton partners. In addition, three scalar multiplets in the fundamental, symmetric two-index, and adjoint representations of SU(3)$_\ell$ provide the symmetry breaking required to reproduce the observed lepton masses and mixings, and allowed gauge boson masses.
The charges and interactions are collected in Table~\ref{tab1} and Eq.~\eqref{eq:Lagrangian}, respectively.

One of the new leptophilic gauge bosons, the one associated with the $L_\mu-L_\tau$ current, is taken to lie in the $10-200$~MeV range, 
being parametrically lighter than the others.
This allows it to address the current discrepancy between experimental measurements and SM predictions of the anomalous magnetic moment of the muon.
Moreover, the Cabibbo angle anomaly can be explained by 
the pair of vector bosons that couple off-diagonally to $e\mu$, with masses in the $1-10$~GeV range, while complying with all other LEP and lepton-flavor universality constraints.

Interestingly, the relative lightness of the $L_\mu-L_\tau$ gauge boson is shown to be linked to a low ($\sim 1$-$100\,$GeV) mass scale for the right-handed neutrinos that participate in the seesaw mechanism.
As shown schematically in Fig.~\ref{fig:mass_hierarchy}, the masses of the HNL eigenstates are inversely proportional to the light neutrino ones, and their mixings with the active flavors match the flavor composition of the corresponding light neutrino mass eigenstates.
These two connections lead us to predict that the lightest active neutrino cannot be arbitrarily light, but must rather have a mass larger than $\sim 0.1\,$~meV.

Once fit to the observed lepton and neutrino properties
(including $(g-2)_\mu$), the gauged SU(3)$_\ell$ scenario is highly predictive and makes potentially testable forecasts for existing and upcoming astrophysical and particle physics experiments.
Focusing on the HNL sector, there are three main avenues that complement each other in testing the parameter space in Fig.~\ref{fig:HNL_yN_yLN}: (\emph{i}) more precise experimental measurements and SM predictions for the muon anomalous magnetic moment;
(\emph{ii}) refined measurements and theoretical calculations of the light-element yields and the number of relativistic degrees of freedom at Big Bang Nucleosynthesis, and (\emph{iii}) improved determinations of the absolute scale of neutrino masses from cosmology or terrestrial experiments like KATRIN~\cite{Aker:2021gma}.
In the charged lepton sector, electroweak precision constraints on lepton mixing with heavy vectorlike partners, needed for realistic lepton mass generation, constitute a powerful test of the model.
These lepton partners might be discoverable at the High-Luminosity LHC or other future colliders. 
In contrast, although they are energetically accessible, the new scalar fields in the model are weakly coupled to the SM states, making their possible detection difficult.

Although it is not mandatory for the model, it is possible to introduce an additional fermionic vectorlike triplet $\Psi_i$ as a dark matter candidate. 
As in the $U(1)_{\L_\mu-\L_\tau}$ model of Refs.~\cite{Foldenauer:2018zrz, Holst:2021lzm}, the observed relic density could be achieved through resonantly enhanced
$\Psi_{\mu/\tau}\bar\Psi_{\mu/\tau}\to \mu^+\mu^- + \nu_{\mu,\tau}\bar\nu_{\mu,\tau}$ annihilations.
The obstacle in the SU(3)$_\ell$ model is that the $\Psi_e$ component would not be sufficiently
suppressed unless its mass also happens to 
allow for resonantly enhanced (co)annihilations through the exchange of the heavier gauge bosons. 
This would require additional model building, which we do not pursue here.

 A remaining challenge for future study is to construct a suitable scalar potential leading to the required $\Phi_3$, $\Phi_6$, and $\Phi_8$ VEVs. Although the preliminary study in Appendix~\ref{app:VEVs} makes it plausible that the desired symmetry breaking pattern can be achieved, further investigation is required 
 to prove it.

It is in principle possible to construct a similar model for quark masses and mixing based on a spontaneously broken SU(3)$_B$ symmetry.
This extension, combined with the present proposal including mixing between the SU(3)$_\ell$ and SU(3)$_B$ gauge bosons, might explain the accumulating evidence for lepton-universality violation in $b\rightarrow s\ell^+\ell^-$ decays~\cite{Geng:2021nhg}.
Work in this direction is underway.

\newpage
{\bf Acknowledgments.}  We thank B.\ Grinstein, A.\ Crivellin, W.\ Altmannshofer, and F. Kahlhoefer for helpful correspondence.
This work was supported by NSERC (Natural Sciences
and Engineering Research Council, Canada).
G.A. is supported by the McGill Space Institute through a McGill Trottier Chair Astrophysics Postdoctoral Fellowship.

\bibliographystyle{utphys}
\bibliography{ref}

\appendix
\section{Scalar interactions}
\label{app:VEVs}
Obtaining the desired pattern of symmetry breaking in a self-consistent way from a scalar potential is a challenging task which is beyond the scope of the present work. 
That said, in this appendix we take a preliminary step in this direction by writing the renormalizable couplings in the scalar sector that are consistent with the gauge symmetries.

The Higgs-like symmetry breaking terms in the scalar potential for the three SU(3)$_\ell$ multiplets have the form
\bea
    V &\ni& \lambda_3 (\Phi^\dagger_3\Phi_3 - v_3^2)^2
    +\lambda_6 (\tr\,\Phi^\dagger_6\Phi_6 - v_6^2)^2\nn\\
    &+&\lambda_8(\tr\,\Phi_8^2 - v_8^2)^2\,.
\eea
Beyond these, the dimensionless interaction terms are
\bea
    V &\ni& \lambda_{3,8}\Phi_3^\dagger\Phi_8\Phi_8\Phi_3 +  \lambda_{3,6}\Phi_3^\dagger\Phi_6^\dagger\Phi_6\Phi_3\nn\\
        &+&\lambda_{6,6}\,\tr(\Phi_6^\dagger\Phi_6\Phi_6^\dagger\Phi_6)
        +\lambda_{8,8}\,\tr(\Phi_8^4)\nn\\
        &+&\lambda_{6,8}\,\tr(\Phi_8^2\Phi_6^\dagger\Phi_6) + \lambda'_{6,8}\,\tr(\Phi_8\Phi_6^\dagger\Phi_8^\T\Phi_6)\nn\\
        &+& \lambda_{3,8}''\,\Phi_3^\dagger\Phi_3\,\tr(\Phi_8^2) + \lambda_{3,6}''\,\Phi_3^\dagger\Phi_3\,\tr(\Phi^\dagger_6\Phi_6)\nn\\
        &+& \lambda_{6,8}''\,\tr(\Phi_8^2)\,\tr(\Phi^\dagger_6\Phi_6)\,,
\eea
while the dimension-3 couplings correspond to 
\bea
    V &\ni& \mu_{3,6}\,\Phi_3^\T\Phi_6\Phi_3 + {\rm H.c.}\nn\\
    &+& \mu_{3,8}\,\Phi_3^\dagger\Phi_8\Phi_3
    + \mu_8\,\tr(\Phi_8^3)\nn\\
    &+& \mu_{6,8}\,\tr(\Phi_6 \Phi_8\Phi_6^\dagger)
    + \mu'_{6,8}\,\tr(\Phi_6\Phi_6^\dagger\Phi_8^\T)\,.
\eea

As a first step, we can set $\Phi_6=0$, $\Phi_3 = (0,0,a)^\T$, and $\Phi_8 = b\lambda_3 + c\lambda_8$ to determine whether the desired values of $a,b,c$ can give rise to a stationary point in the potential within this restricted slice of the field space. 
There are 9 relevant parameters in the potential, three equations of stationarity for given values of the VEVs, and three stability conditions, so this seems generically possible.

The more difficult step may be to arrive at the desired form of $\langle\Phi_6\rangle$. 
In a linear analysis, due to the source term $\mu_{3,6}\Phi_{3,i}\Phi_{3,j}$, one would expect the $\Phi_6$ VEV to be dominantly aligned in the $(3,3)$ direction.
This is in contrast with the need for relatively democratic entries in order to recover the observed mixing pattern of active neutrinos as dictated by the PMNS matrix. 
One should then suppress the $\mu_{3,6}$ coupling and obtain the desired solution at the nonlinear level of the minimization equation.
This would yield the schematic form
\be
    A\Phi_6 + B\Phi_6^3 = \epsilon\,,
\ee
where $\epsilon$ is the suppressed source term.  There are 
 14 real potential parameters involving  $\Phi_6$, compared to 6 extremization and 6 stability conditions; therefore it may be possible to obtain the desired pattern of $\Phi_6$ VEVs.
We defer a more thorough analysis  to future study.

\end{document}